\documentclass[a4paper,twocolumn,showpacs,superscriptaddress,floatfix]{revtex4-2}

\usepackage{graphicx}
\usepackage{epsf}
\usepackage{epsfig}
\usepackage{amssymb,amsmath}
\usepackage[usenames]{color}
\usepackage{amssymb}
\usepackage{times}
\usepackage{tabularx}
\usepackage[table]{xcolor}

\newcommand{\be}{\begin{equation}}
\newcommand{\ee}{\end{equation}}
\newcommand{\bea}{\begin{eqnarray}}
\newcommand{\eea}{\end{eqnarray}}
\newcommand{\nn}{\nonumber}

\font\tenscr=rsfs10 scaled1100
\font\sevenscr=rsfs7 % scaled \magstep1
\font\fivescr=rsfs5 % scaled \magstep1
\skewchar\tenscr='177
\skewchar\sevenscr='177
\skewchar\fivescr='177
\newfam\scrfam
\textfont\scrfam=\tenscr
\scriptfont\scrfam=\sevenscr
\scriptscriptfont\scrfam=\fivescr

\begin{document}

\title{A Weyl's law for black holes}

\author{Jos\'e Luis Jaramillo}
\affiliation{Institut de Math\'ematiques de Bourgogne (IMB), UMR 5584, CNRS, \\ Universit\'e de
Bourgogne, F-21000 Dijon, France}
\author{Rodrigo Panosso Macedo}
\affiliation{Niels Bohr International Academy, Niels Bohr Institute, Blegdamsvej 17, 2100 Copenhagen, Denmark}
\affiliation{School of Mathematical Sciences and STAG Research Centre, University of Southampton, Southampton, SO17 1BJ, United Kingdom}
\author{Oscar Meneses-Rojas}
\affiliation{Institut de Math\'ematiques de Bourgogne (IMB), UMR 5584, CNRS, \\ Universit\'e de
Bourgogne, F-21000 Dijon, France}
\author{Bernard Raffaelli}
\affiliation{Institut de Math\'ematiques de Bourgogne (IMB), UMR 5584, CNRS, \\ Universit\'e de
Bourgogne, F-21000 Dijon, France}
\author{Lamis Al Sheikh}
\affiliation{Institut de Math\'ematiques de Bourgogne (IMB), UMR 5584, CNRS, \\ Universit\'e de
Bourgogne, F-21000 Dijon, France}
\affiliation{Soci\'et\'e ``gayte.it'', Si\`ege - 46, Le Villeret, 48170 St. Jean La Fouillouse, France}

\begin{abstract}
  We discuss a Weyl's law for the quasi-normal modes of black holes
  that recovers the structural features of the standard Weyl's law
  for the eigenvalues of Laplacian-like operators in compact regions.
  Specifically, we propose that the asymptotics of the counting function $N(\omega)$
  of quasi-normal modes of $(d+1)$-dimensional
  black holes follows a power-law $N(\omega)\sim \mathrm{Vol}_d^{\mathrm{eff}}\omega^d$,
  with $\mathrm{Vol}_d^{\mathrm{eff}}$ an effective $d$-volume determined by the
  light-trapping properties of the black hole geometry. Concretely, the factorisation
  $\mathrm{Vol}_d^{\mathrm{eff}} \sim \left(8\pi/\kappa\right) \cdot \mathrm{Vol}^{\mathrm{trapped}}_{d-1}$
  makes apparent the two underlying structural ingredients, namely the (local) redshift effect
  controlled by the surface gravity $\kappa$ and the volume $\mathrm{Vol}^{\mathrm{trapped}}_{d-1}$ of the
  (phase space) trapped set. In particular, this proposal extends the Weyl's law proved by
  Dyatlov \& Zworski for the counting of slowest decaying quasi-normal modes, to include overtones. 
  As an application, these  Weyl's laws could provide a probe into the
  effective spacetime dimensionality, upon the counting of sufficiently many
  quasi-normal modes in the ringdown signal of binary black hole mergers.

\end{abstract}


\maketitle

\section{Spectrum asymptotics, degrees of freedom counting and geometry}
Our goal is to motivate and propose a
Weyl's law for black hole (BH) quasi-normal modes (QNMs).
Such Weyl's law provides an estimate of the number of QNM frequencies, 
contained inside a circle  of a given radius in the frequency-complex plane,
in terms of the asymptotics of
BH QNM overtones and a particular (volume) scale defined by the geometry of the BH spacetime.
It directly extends to the BH QNM setting the standard Weyl's law
for the counting of  eigenvalues of the Laplacian defined on a compact Riemannian manifold.

\subsection{Weyl's law for normal modes: selfadjoint case in a nutshell}
Weyl's law~\cite{Weyl11,Weyl12} provides a landmark in spectral geometry.
From a physical perspective, it relates three distinct notions:
the number of states of a physical system contained in a given spatial domain $D$,
the high-frequency asymptotics of the spectrum of the operator controlling its dynamics
(the time infinitesimal generator), and the geometry of the region $D$. 
The study of this triple relation has proved to be a far-reaching catalyst for key
developments both in physics and mathematics.
On the physics side, it lays e.g. at the core of the developments in radiation theory
aiming at understanding the black body radiation and, therefore, it is ultimately interwoven
with the birth of quantum theory (see e.g. \cite{Kac66,AreNitPet09}). On the mathematical
side, it dwells in the more general setting of inverse problems in spectral geometry that have
unveiled a rich set of correspondences between the  spectral properties of geometric
differential operators defined on (Riemannian) manifolds, namely the Laplacian,
and the properties of families of geodesics on such manifolds  
\cite{Kac66,Berger03,Ivrii16}.
Specifically, Weyl's law for the Laplacian
stands as an archetype of such results in spectral geometry with direct
implications in physics.

For concreteness, let us consider the  asymptotics of the homogenous
scalar Helmholtz equation on a compact domain $D$ 
\bea
\label{e:Helmhotz}
(\Delta + k_n^2)\phi_n  = 0 \ ,
\eea
subject to homogeneous (Dirichlet, Neumann or Robin)
boundary conditions on its boundary $\partial D$. From the compacity of $D$ the 
wave-number's norm eigenvalues $k_n^2$ form a discrete set and, denoting by
$N(k)$ the number of $k_n$'s (positive square-root of actual eigenvalues $k_n^2$)
below a certain wave-number $k$,
the following large-$k$ asymptotic relation holds
\bea
\label{e:Weyl_Helmhotz}
N(k) \sim \mathrm{Vol}_d(D) k^d +  o(k^{d-1}) \ \ , \ \ (k\to \infty)
 \eea
 where $\mathrm{Vol}_d(D)$ is the volume of the compact domain $D$ (e.g.
 of the optical cavity in the black body problem) and $d$ is
the space dimension of $D$, i.e. $d=\mathrm{dim}(D)$. This is the Weyl's law.
Crucially, this `universal' asymptotic leading term does not depend on the shape of
$D$, in particular the boundary $\partial D$ only entering at the subleading term in the
asymptotic expansion (see e.g. \cite{BaltesHilf}).
Similar Weyl's law asymptotics hold for other fields (such as the electromagnetic field
in the  black body problem)
and other selfadjoint differential operators \cite{Ivrii16} for the counting of the corresponding
(discrete) normal modes, the latter characterised by the spectral theorem.

\subsection{An invitation to Weyl's law for BH QNMs}
\label{s:Invitation_BH_QNM_Weyl}
In ref. \cite{Jaramillo:2021tmt} a Weyl's law was proposed for BH QNMs.
QNMs encode the linear response of a resonator under
an external perturbation, corresponding to exponentially decaying
oscillating solutions to an appropriate wave equation under outgoing boundary conditions
in a (generally) non-compact domain $D$ (cf.
\cite{Kokkotas:1999bd,Nollert:1999ji,Berti:2009kk,Konoplya:2011qq} for reviews of QNMs in the
gravitational setting and \cite{LalYanVyn17} for optical nanoresonators).
Specifically in the BH case, QNM complex frequencies $\omega_n$ (the real part
providing the QNM oscillation frequency and the imaginary part the inverse of its time-decay
scale) encode invariant information on the BH stationary background. A crucial feature
of QNM frequencies, in spite of the non-compactness of $D$, is that they form a
discrete set. The number of frequencies $\omega_n$ in a given circle of the
complex $\omega$-plane can therefore be counted and the question about a possible QNM version of
Weyl's law is meaningfully posed. Consequently, Weyl's laws for QNMs have been the subject of
extensive research
(see e.g. \cite{Zworski99,PotWeiBar12,Sjost14,zworski2017mathematical,dyatlov2019mathematical}).
    However, the situation for QNMs (and, likewise and closely related, for non-selfadjoint operators
    where the spectral theorem is absent; cf.~\cite{Sjostrand2019} and references therein) 
    is more subtle than in the standard
selfadjoint case in Eq. (\ref{e:Helmhotz}). In particular, a Weyl's law analogous
to relation (\ref{e:Weyl_Helmhotz}) is not generic
beyond the one-dimensional case and for compact support (or very fastly
decaying) scattering potentials~\cite{Zwors87,Froes97}.

Specially relevant in our BH setting are the groundbreaking results by
Dyatlov~\cite{Dyatlov:2013hba} and Dyatlov \& Zworski~\cite{Dyatlov:2013fua}
in the Kerr-de Sitter context
(see also related results in other open systems \cite{PotWeiBar12,BarWeiPot13}).
In these articles a BH QNM Weyl's law for BH spacetimes is formulated and proved.
Such Weyl's law provides a counting of (a subset of) the BH QNM frequencies in the spirit of
the asymptotics in relation  (\ref{e:Weyl_Helmhotz}). Specifically, for $(3+1)$-Kerr-de Sitter
spacetimes, the BH QNM Weyl's law in Refs.~\cite{Dyatlov:2013hba,Dyatlov:2013fua} provides
(cf. Eq. (2) in~\cite{Dyatlov:2013fua} and Theorem 3 in~\cite{Dyatlov:2013hba})
the asymptotics for a QNM counting function $N_{_{\mathrm{DZ}}}(\omega)$ (in terms of the frequency
$\omega$, better suited to QNMs than the wave-number $k$) in the following form
\bea
\label{e:Weyl_Dyatlov-Zworski}
N_{_{\mathrm{DZ}}}(\omega)\sim C \; \omega^2 \ ,
\eea
(the precise definition of $N_{_{\mathrm{DZ}}}(\omega)$ will be introduced later in
    Eq. (\ref{e:counting_function_QNMs_DZ}).
The constant $C$ is determined by the (symplectic) volume of the 
{\em trapped set}, namely the set (in phase space) determined by bounded
null geodesics, i.e. not traversing the BH horizon or escaping to infinity
(see discussion and references in~\cite{Dyatlov:2013hba,Dyatlov:2013fua}).
The BH QNM Weyl's law (\ref{e:Weyl_Dyatlov-Zworski}) differs from the
selfadjoint expression (\ref{e:Weyl_Helmhotz})
in two important aspects: i) in spite of working in a $3$-dimensional spatial setting, the power $2$ in the BH QNM case (\ref{e:Weyl_Dyatlov-Zworski}) differs from the power
$3$ that would be consistent with $d=3$ in (\ref{e:Weyl_Helmhotz}); ii) the constant
$C$ is not directly related to a $3$-dimensional ``region/cavity'' in spatial
(Riemannian) sections of the BH geometry but rather to the volume of time constant sections of
the trapped region in the phase space (symplectic) geometry, actually of dimension $4$
(cf.~\cite{Dyatlov:2013hba,Dyatlov:2013fua} and section \ref{PhaseSpace_TrappedSet}).
Note however that points i) and ii) are indeed consistent,
in the sense that the corresponding ``space''-dimension of the trapped region, i.e. $1/2\cdot 4= 2$, is indeed
the exponent of the power-law in the BH QNM Weyl's law.

In the light of the previous discussion, it seems remarkable
that a BH QNM Weyl's law can actually be (experimentally) found~\cite{Jaramillo:2021tmt} that
`faithfully' corresponds to the selfadjoint case asymptotics in expression (\ref{e:Weyl_Helmhotz}). Namely,
for a $(3+1)$-dimensional BH spacetime a Weyl's power law is found to follow
  $N\sim \omega^3$, with the `correct' exponent 
  corresponding to the space dimension.
In the present work we justify and extend such claims in~\cite{Jaramillo:2021tmt}
for the QNM Weyl's law of stationary black holes,
in particular providing the details of the role played by 
trapping in the identification of the exponent in Weyl's law power-law.
In a further key step, the present analysis identifies the geometric structure of the
multiplicative constant in BH QNM Weyl's law (a discussion completely absent in~\cite{Jaramillo:2021tmt}).
Specifically, such multiplicative constant is given by an effective volume $\mathrm{Vol}_d^{\mathrm{eff}}$ 
that is accounted solely in terms of the BH (semi-Riemannian)
spacetime geometry, namely by null geodesic trapping. 
As a result, considering a  $(d+1)$-dimensional BH spacetime, we find
support~\footnote{This claim is independent of the spacetime asymptotics at infinity. Although
the current discussion dwells mostly in the asymptotically flat (spherically
symmetric) BH setting, QNM universality results
in~\cite{Jaramillo:2021tmt,Gasperin:2021kfv,Destounis:2021lum,boyanov2023pseudospectrum}, together with ongoing
work~\cite{Besson_et_al},
support its validity in the generic BH case, including asymptotically flat,
de Sitter and anti-de Sitter BH spacetimes.} for
\bea
\label{e:BH_QNM_Weyl}
N(\omega) \sim \frac{1}{c^d}\mathrm{Vol}_d^{\mathrm{eff}} \omega^d +  o(\omega^{d-1}) \ \ , \ \ (\omega\to \infty)
\eea
where $N(\omega)$ amounts for the number of (complex) QNM frequencies $\omega_n$
such that $|\omega_n|\leq \omega$, where $\omega>0$ (the appropriate
  power of the light speed $c$ has been introduced for dimensional reasons).
  As commented above, the  constant  $\mathrm{Vol}_d^{\mathrm{eff}}$ is controlled
  by trapping and, more specifically, such  effective volume  can be 
factorised into two terms: a first (radial) factor determined by a (local) redshift effect
---and controlled 
by the surface gravity $\kappa$--- and a second (angular) factor given by the
(phase space) trapped set (namely $C$ in Eq. (\ref{e:Weyl_Dyatlov-Zworski}) above).
The explicit multiplicative structure of the $\mathrm{Vol}_d^{\mathrm{eff}}$
$d$-dimensional effective volume is presented below in (\ref{e:BH_QNM_Weyl_dim-d}).

The present work provides a reconcialiation between the results
in~\cite{Jaramillo:2021tmt} and~\cite{Dyatlov:2013hba,Dyatlov:2013fua}.
In particular,  Weyl's law~(\ref{e:BH_QNM_Weyl}) is not in contradiction with Dyatlov \& Zworski Weyl's
law~(\ref{e:Weyl_Dyatlov-Zworski}) since the respective counting
functions $N(\omega)$ and $N_{_{\mathrm{DZ}}}(\omega)$ differ in their precise definition
(see respective Eqs. (\ref{e:counting_function_QNMs}) and (\ref{e:counting_function_QNMs_DZ}) below). 
Weyl's law~(\ref{e:BH_QNM_Weyl}) extends and reconcialiates the results
in ~\cite{Jaramillo:2021tmt} and~\cite{Dyatlov:2013hba,Dyatlov:2013fua}, by complementing the former~\footnote{Ref.~\cite{Jaramillo:2021tmt} discusses the structural stability of the BH QNM Weyl's law
    under background ultraviolet perturbations, that extends the commented Weyl law's robustness under
    spacetime asymptotics.
    Here we make the methodological choice of focusing on the non-perturbed (stationary)
    BH case, leaving the perturbed case for a devoted discussion~\cite{Besson_et_al} (see also
    section~\ref{s:Weyl_perturbed_BHs}).}
with the identification of the multiplicative constant and extending
the latter with the counting of all BH QNMs, that renders the space dimension $d$ as the
exponent in the power law~\footnote{We note that the status of Weyl's law (\ref{e:Weyl_Dyatlov-Zworski}) is in stark
contrast with that of version~(\ref{e:BH_QNM_Weyl}) of Weyl's law: whereas Dyatlov \& Zworski's
law (\ref{e:Weyl_Dyatlov-Zworski}) is a theorem with a complete proof (Theorem 3 in~\cite{Dyatlov:2013hba}),
the present version ~(\ref{e:BH_QNM_Weyl}) stands as a heuristic proposal lying on
semi-analytical and numerical support.}.

\subsubsection{BH QNM versus selfadjoint Weyl's law: an interpretation shift}

Let us consider more closely the comparison between the proposed BH QNM Weyl's law (\ref{e:BH_QNM_Weyl})
and the selfadjoint Weyl's law  (\ref{e:Weyl_Helmhotz}). Regarding the exponent in the power law,
the agreement is apparent since its value is given by $d$ in both cases.
However regarding $\mathrm{Vol}_d^{\mathrm{eff}}$, given the non-compact support of the BH geometry,
this multiplicative constant does not admit a straightforward interpretation as a compact
``cavity'' volume, as for $\mathrm{Vol}_d(D)$ in Weyl's law (\ref{e:Weyl_Helmhotz}). Rather,
in the strategy we have followd above, it is natural
to seek a (spacetime) interpretation of  $\mathrm{Vol}_d^{\mathrm{eff}}$ as providing an
`effective', rather than an `actual', $d$-dimensional scale where the resonance phenomenon takes place.
In this spirit, a key step in our reasoning below consists in shifting
the interpretation of $\mathrm{Vol}_d^{\mathrm{eff}}$ from a $(\mathrm{Lenght})^d$ $d$-volume to rather
a $(\mathrm{Time}\cdot\mathrm{Area}_{d-1})$ $d$-volume. This can be seen as amounting to a shift
in the perspective on $N(\omega)$, from its role in defining a `density of states'
$\rho(\omega) \sim N(\omega)/\mathrm{Vol}_d^{\mathrm{eff}}$ to rather
capturing a notion of `radiation flux' $F(\omega)\sim N(\omega)/(\mathrm{Time}\cdot\mathrm{Area}_{d-1})$.
In this setting, the `angular resonant volume' $\mathrm{Area}_{d-1}$ is characterised by
the notion of `trapped region' volume (in phase space) in Dyatlov \& Zworski
Weyl's law~(\ref{e:Weyl_Dyatlov-Zworski}), whereas the $\mathrm{Time}$ (radial) term is provided
by a `dissipation time' controlled by the BH surface gravity $\kappa$. As we will see
below,  this leads to a surprisingly
compact and geometric expression for $\mathrm{Vol}_d^{\mathrm{eff}}$.

With these elements, we can give now a taste of the use of BH QNM counting asymptotics that we have in mind.
Specifically, we aim at inverting the logic in reading BH QNM Weyl's law (\ref{e:BH_QNM_Weyl}), as compared
to that of the standard selfadjoint Weyl's law (\ref{e:Weyl_Helmhotz}). Whereas in the latter the
dimension $d$ and the cavity's volume $\mathrm{Vol}_d(D)$ are known~\footnote{Alternatively, the interest may
  lie only on the density of states $\rho(\omega)=N(\omega)/\mathrm{Vol}_d(D)$, as in the black
  body radiation problem.
  In either case, the cavity's volume $\mathrm{Vol}_d(D)$ is not the quantity of interest, 
  just playing the (key) role of being the only trace of the cavity's geometry and guaranteeing
  Weyl's law universality (actually the original motivation for Weyl's proof).} a priori, in
  the former BH QNM case they can be taken as unknown. In this setting,
  the BH QNM Weyl's law (\ref{e:BH_QNM_Weyl}) can be used to probe:
\begin{itemize}
\item[i)] The (effective) spacetime dimension $(d+1)$.
\item[ii)] The dynamical scale for the BH resonant phenomenon,
\end{itemize}
under the assumption that $N(\omega)$ can be independently (e.g. observationally) accessed.
The possible use of the proposed BH QNM Weyl's law as a probe into spacetime features defines our
interest in this QNM asymptotic counting problem.

\medskip

The article is organized as follows.
In section \ref{s:Weyl_law} we review
the Weyl's law in the classical setting of the eigenvalues of
the Laplacian in a compact region. In section \ref{s:QNM_Weyl} we discuss
the Weyl's law for QNMs associated with the scattering (in non-compact domains)
by compact support potentials, separating the special one-dimensional case
from the general discussion in higher dimensions. In section \ref{s:QNM_Weyl_BHs}
we discuss the application to the BH case, focusing on
the Schwarzschild BH case and then proposing a QNM Weyl's law for generic BHs.
Finally, section \ref{s:conclusions} presents the conclusions and
some perspectives.

\section{Weyl's law: space(-time) dimension and length scales}
\label{s:Weyl_law}
We present here a brief account of the elements of Weyl's law
relevant in our later discussion on BH QNMs, with a focus on the volume factor
and the power-law dependence on the dimension $d$. We review first the Weyl's law in the classical selfadjoint
case and, in a second stage, section \ref{s:QNM_Weyl} presents some results
in the QNM (scattering resonances) setting.

  \subsection{Weyl's law in closed manifolds}
  For concreteness, we focus on the case of a compact manifold
  without boundaries~\cite{Berger03}. This case contains all the specific elements we aim
  at addressing in the QNM case, providing a simplified, but adequate setting.
  The  same leading-order result is found for compact manifolds with boundary,
  the differences related to the boundary terms only appearing at
  the subleading-order, that we will not address here.

Given a closed (compact without boundary) Riemannian manifold
$(D, h_{ab})$ of dimension $d$ with Levi-Civita connection
$\nabla_a$, we consider the eigenvalue problem associated with
scalar Laplace-Beltrami operator
$\Delta_h = h^{ab}\nabla_a\nabla_b$, namely 
\bea
\label{e:Laplace_closed}
-\Delta_h\phi_n = \lambda_n\phi_n \ .
\eea
The eigenvalues $\lambda_n$ are real and non-negative. 
Given $\lambda\in\mathbb{R}^+$ we define the eigenvalue counting function
$N(\lambda)$ as
\bea
N(\lambda) = \#\{ \lambda_n\in\mathbb{R}, \hbox{ such that } \lambda_n\leq \lambda\} \ .
\eea
The Weyl's law states that, in the large-$\lambda$ asymptotic limit 
\bea
\label{e:counting_function_compact}
N(\lambda) \sim C_d \mathrm{Vol}_d(D) \lambda^{\frac{d}{2}} \ \ , \ \ (\lambda\to\infty)
 \eea
where $\mathrm{Vol}_d(D)$  is the $d$-dimensional volume of $D$ and 
$C_d$ are universal constants only depending on the dimension $d$
\bea
C_d = \frac{\mathrm{Vol}_d(B^d_1)}{(2\pi)^d} \ ,
\eea
with $\mathrm{Vol}_d(B^d_1)$ the Euclidean volume of the unit $d$-dimensional 
ball $B^d_1$. To get a taste and for later comparison,
we give some examples for the lowest dimensions:
\begin{itemize}
\item[i)] {\em Case $d=1$}: for the ball $B^1_1$, namely the $[-1,1]$ interval
  \bea
  \mathrm{Vol}_1(B^1_1) = 2 \ \ , 
  \eea
  and therefore, denoting $\mathrm{Vol}_1(D) = L$, we get
  \bea
  \label{e:Weyl_law_onedim}
  N(\lambda) \sim \frac{2}{2\pi}L \lambda^{\frac{1}{2}} =\Big(\frac{L}{\pi}\Big) \lambda^{\frac{1}{2}}
\ \ , \ \ (\lambda\to\infty) \ .
  \eea

\item[ii)] {\em Case $d=2$}: for the ball $B^2_1$, the unit circle in the plane%, gives
 \bea
  \mathrm{Vol}_2(B^2_1) = \pi \ \ , 
  \eea
  and denoting $\mathrm{Vol}_2(D) = A$, namely the area, we get
  \bea
  N(\lambda) \sim \frac{\pi}{(2\pi)^2} A \; \lambda = \Big(\frac{A}{4\pi}\Big)\lambda 
\ \ , \ \ (\lambda\to\infty) \ . 
  \eea
  
\item[iii)] {\em Case $d=3$}:  the ball $B^3_1$ is the unit Euclidean ball, so 
 \bea
  V_3(B^2_1) = \frac{4}{3}\pi \ \ , 
  \eea
  and denoting $\mathrm{Vol}_3(D) = V$ we get
  \bea
  \label{e:C_3}
  N(\lambda) \sim \frac{4\pi/3}{(2\pi)^3} V \; \lambda^{\frac{3}{2}}
  = \Big(\frac{V}{6\pi^2}\Big)\lambda^{\frac{3}{2}} \ \ , \ \ (\lambda\to\infty) \ .
  \eea

\item[iv)] {\em Case general $d$}: for dimension $d$, the unit Euclidean $d$-ball has
  volume (with $\Gamma(x)$ Euler's gamma function)
\bea
  \mathrm{Vol}_d(B^d_1) = \frac{\pi^{\frac{d}{2}}}{\Gamma(\frac{d}{2}+1)} \ \ , 
  \eea
  and denoting $\mathrm{Vol}_d(D) = V_d$ we get
 \bea
  N(\lambda) &\sim& \frac{\pi^{\frac{d}{2}}/\Gamma(\frac{d}{2}+1)}{(2\pi)^d} V_d \lambda^{\frac{d}{2}} \nn \\
  &=& \bigg(\frac{V_d}{(4\pi)^{\frac{d}{2}}\Gamma(\frac{d}{2}+1)}\bigg)\lambda^{\frac{d}{2}}
  \ \ , \ \ (\lambda\to\infty) \ .
  \eea

\end{itemize}
The important point we would like to stress here is that, from the leading
order of Weyl's law, one can extract both the dimension of the manifold
and an averaged (volumetric) typical length scale $L_{\mathrm{vol}}:=(V_d)^{\frac{1}{d}}$.

\subsection{Weyl's law in compact manifolds with boundaries}
As commented above, the Weyl's law asymptotics for the leading term of the eigenvalue
counting function $N(\lambda)$ holds also for the spectral problem of the Laplacian in
a compact domain $D\subset \mathbb{R}^d$ with boundaries. The differences in $N(\lambda)$
following from distinct boundary conditions show up only at the
subleading order. Let us consider the spectral problem
\bea
\label{e:Laplacian_compact_boundaries}
\left\{
\begin{array}{lcl}
-\Delta \phi_n = \lambda_n \phi_n \ \ &,& \ \ (D\subset  \mathbb{R}^d, D \hbox{ compact}) \\
\left.\phi_n\right|_{\partial D} = 0 \ \ &,& \ \ \hbox{(Dirichlet boundary conditions)}\\
\hbox{or} &&\\
 \left.\partial_s\phi_n\right|_{\partial D} = 0  \ \ &,& \ \ \hbox{(Neumann boundary conditions)}
\end{array}
  \right.  \ ,\nonumber\\
\eea
with $\partial_s\phi_n$ the normal derivative at the boundary $\partial D$.
In this case, under appropriate conditions~\cite{Ivrii16}, the `Weyl's conjecture' 
\bea
N(\lambda) \sim C_d \mathrm{Vol}_d(D) \lambda^{\frac{d}{2}} \mp \frac{1}{4}
 C_{d-1} \mathrm{Vol}_{d-1}(\partial D) \lambda^{\frac{d-1}{2}} \ \ , 
 \eea
 holds when $\lambda\to\infty$, with the ``minus'' sign corresponding to the Dirichlet case and the ``plus'' sign
 to the Neumann one.

\section{QNM Weyl's law}
\label{s:QNM_Weyl}
For the sake of concreteness in the introduction of QNMs, we consider
the scattering problem corresponding to
the wave equation of a scalar field propagating
in the (non-compact) space $\mathbb{R}^d$ (namely scalar wave in $(d+1)$-dimensional
Minkowski spacetime) with scattering potential $V(x)$, where $x\in \mathbb{R}^d$
and $V(x)$ is time-independent~\footnote{More general time-dependent situations
  can be considered (see e.g.~\cite{SofWei98}), but in the present Weyl's law
  setting we keep ourselves in the time-independent case.}
\bea
\label{e:wave_equation_d-dims}
\left(\partial^2_t - \Delta + V(x)\right) \phi= 0 \ ,
\eea
subject to outgoing boundary conditions and with appropriate
initial data. By taking the Fourier transform in time, and
denoting the frequency spectral parameter as $\omega$, we can write
the following spectral problem
\bea
\label{e:QNM_equation}
\left(- \Delta + V(x)\right) \phi= \omega^2 \phi \ , 
\eea
complemented with the asymptotic outgoing boundary conditions.
Introducing the Schr\"odinger operator
\bea
\label{e:P_operator}
P = -\Delta + V(x) \ , 
\eea
QNMs can be understood ``heuristically''
as corresponding to the discrete part
~\footnote{In this heuristic perspective, the continuous
  part of the spectrum would correspond to the so-called `tails'.}
of the spectrum of this
``operator'', where the role of outgoing boundary conditions is crucial
\bea
\label{e:QNM_wave}
\left\{
  \begin{array}{l}
    P\phi_n = \left(-\Delta + V(x)\right)\phi_n = \omega^2_n \phi_n  \ \ , \ \ \hbox{(in } \mathbb{R}^d)\\
    \hbox{Outgoing boundary conditions at infinity} \ .
  \end{array}
\right.
    \eea
The dissipative character of the outgoing boundary conditions entails
the non-selfadjoint character of the ``operator'', so the QNM
frequencies $\omega_n$'s are generically complex numbers.
From a more formal perspective, QNMs can be introduced in terms of
the resolvent $R_P(\omega) = (P- \omega \mathrm{I})^{-1}$ of $P$.
Under the appropriate conditions on $V(x)$ (and with the appropriate
convention for $\omega$, namely $\phi(t,x)\sim e^{i\omega t}\phi(x)$),
the resolvent $R_P(\omega)$ is analytic
in the lower-half $\omega$-complex plane. QNMs frequencies $\omega_n$  
correspond then to the poles of the meromorphic extension of the
resolvent $R_P(\omega)$ to the upper-half $\omega$-complex plane
(see, for instance \cite{zworski2017mathematical,dyatlov2019mathematical}).
Alternatively, a characterization of QNMs can be given as proper eigenvalues of
an non-selfadjoint operator defined in terms of hyperboloidal spacetime
foliations \cite{Zenginoglu:2011jz,Warnick:2013hba,Ansorg:2016ztf,Bizon:2020qnd,Jaramillo:2020tuu}.
The latter is actually the approach we adopt in the calculations presented below.

A crucial feature in the present Weyl's law setting is the discrete nature
of QNM frequencies $\omega_n$'s, in spite of the non-compact
nature of the integration domain. This allows to introduce a counting
function for such $\omega_n$'s, as in the case of eigenvalues $\lambda_n$'s of
the Laplace operator in compact domains, namely Eqs. (\ref{e:Laplace_closed})
and (\ref{e:Laplacian_compact_boundaries}). However, given that the problem is
now being defined in the complex plane (not an ordered set), different
natural strategies for the counting can be considered.
In this setting, given $\omega\in \mathbb{R}^+$,
we consider here (but crucially not in~\cite{Dyatlov:2013hba,Dyatlov:2013fua})
the QNM counting function $N(\omega)$ defined as
\bea
\label{e:counting_function_QNMs}
N(\omega) = \# \{ \omega_n \in \mathbb{C}, \hbox{ such that } |\omega_n|\leq \omega \} \ ,
\eea
that is, $N(\omega)$ counts the number of complex QNM frequencies $\omega_n$'s
contained in a circle in the complex plane of radius $\omega$ and  centered in the origin.
Having introduced the counting function $N(\omega)$ the study of its
large-$\omega$ asymptotics is naturally posed, in particular the question about a possible Weyl's law.

The assessment of the Weyl's law in the case of QNM frequencies is more
delicate than in the classical selfadjoint case. A complete answer can be
given in the one-dimensional case for compact support or fast decreasing potentials,
but the situation becomes more complicate in higher dimensions~\footnote{In contrast,
a powerful approach in terms of a semi-classical treatment emerges, providing
rich information on the distribution of QNMs in such semiclassical limit (cf. e.g. \cite{Sjost14}).
This is the setting of the BH QNM Weyl's law in~\cite{Dyatlov:2013hba,Dyatlov:2013fua}.}.
In the following we briefly review these respective cases.

\subsection{QNM Weyl's law: one-dimensional case}
\label{s:Weyl_QNM_one-dim}
\begin{figure}[t!]
\centering
\includegraphics[width=8.4cm]{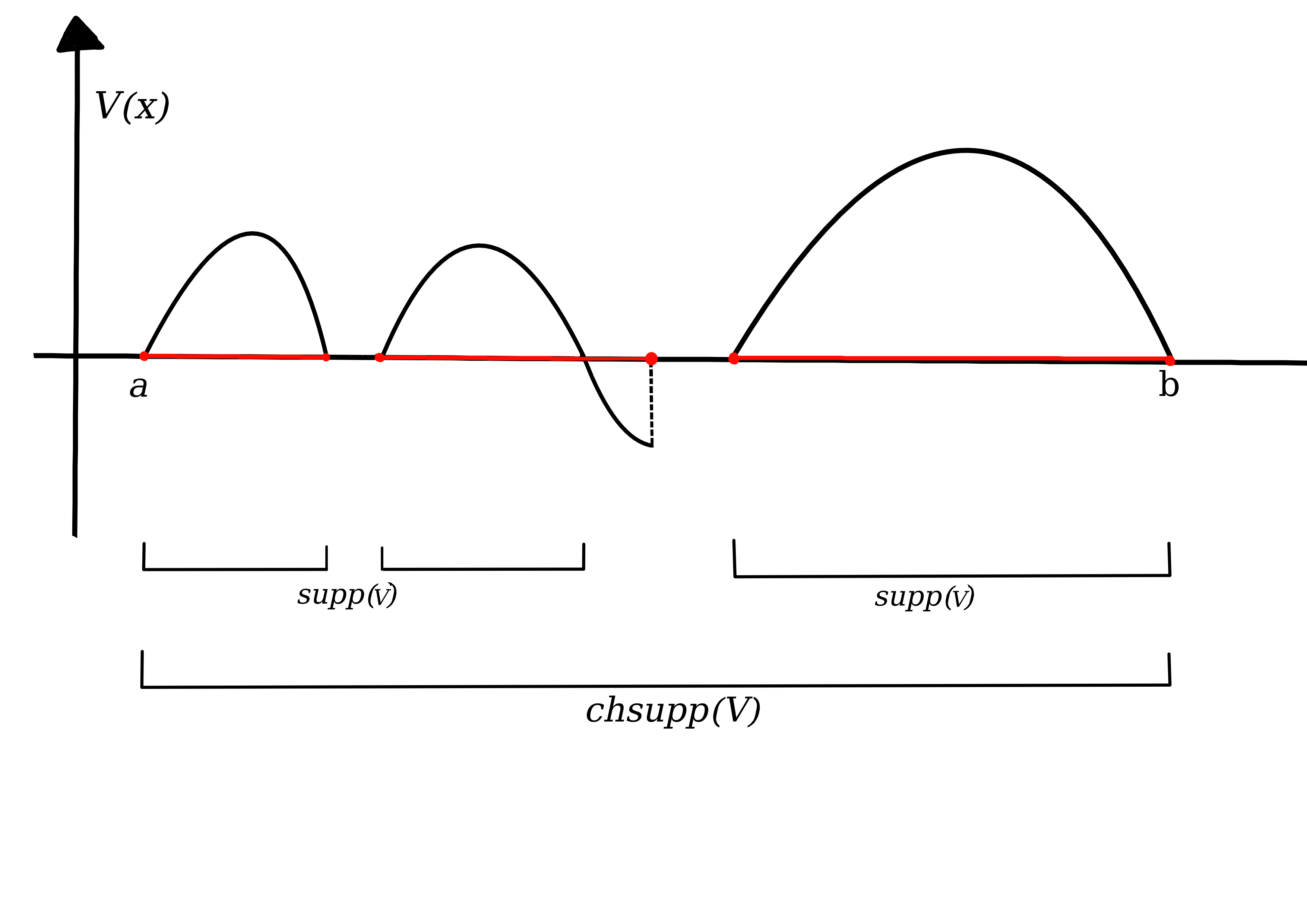}
\caption{
  Illustration of the ``convex hull'' $\mathrm{chsupp(V)}$ of the support of a potential
  $V(x)$. This notion, namely the ``diameter'' of the region where the potential
  does not vanish, provides the relevant multiplicative scale in QNM Weyl's law for compact support potentials.
 }
\label{fig:chsupp}
\end{figure}

Let us consider a compact support, bounded potential $V(x)$ in the one-dimensional case $d=1$.
Given the support of $V(x)$, denoted as $\mathrm{supp(V)}$, the 
so-called ``convex hull'' of $\mathrm{supp(V)}$, respectively denoted as $\mathrm{chsupp(V)}$,
is the minimal convex set containing  $\mathrm{supp(V)}$. In other words,
$\mathrm{chsupp(V)}$ is the smallest interval containing the support of $V(x)$
(cf. Fig.~\ref{fig:chsupp} for an illustration of this concept).
Denoting by $L$ such $\mathrm{chsupp(V)}$, the following asymptotic Weyl's
law in one dimensions holds~\cite{Regge58,Zwors87,zworski2017mathematical,dyatlov2019mathematical}
\bea
\label{e:Weyl_QNM_one-dim}
N(\omega) \sim  2 \left(\frac{L}{c\pi}\right) \omega \ \ , \ \ (\omega\to\infty) \ .
\eea
We note that this expression recovers, up to a factor $2$, Weyl's law (\ref{e:Weyl_law_onedim}),
with the $\omega/c = k = \lambda^{\frac{1}{2}}$ (where $k$ is the
``wave number'' in expression (\ref{e:Weyl_Helmhotz}) and $c$ is the
``light speed'') and with $L$ identified as the ``effective volume''
associated with the potential $V(x)$. This follows from the
existence of two QNM branches, symmetric with respect to the imaginary axis. 
We discuss further this in section \ref{s:therm_time}, reinterpreting $L$.

For completeness (and later convenience), we note that this one-dimensional Weyl's law is intimately
related to the following
result~\cite{Regge58,Zwors87} on the asymptotic distribution of the QNMs frequencies $\omega_n$'s,
with $n\gg 1$, of a one-dimensional potential $V(x)$
of low regularity (class $C^p$, $p<\infty)$
\bea
\label{e:Regge_branches}
\!\!\!\!\!\omega^{\rm R}_n \sim \pm\frac{\pi c}{L}\left( n + \tilde \gamma \right), \,\,
\omega^{\rm I}_n \sim \frac{c}{L} \bigg[\gamma\ln \left( \left|\omega_n^{\rm R}\right| + \gamma'\right) - \ln S\bigg],\nonumber\\
\eea
where $\gamma$, $\tilde{\gamma}$, $\gamma'$ and $S$ are constants depending on the potential $V(x)$.
This asymptotic distribution of $\omega_n$'s defines the so-called ``Regge QNM branches''~\cite{Regge58,Zwors87}. 
Note the asymptotic regular distribution of frequencies $\omega_n$'s along the logarithmic QNM
branches, with constant $\Delta \omega^{\rm R} := \omega^{\rm R}_{n+1}-\omega^{\rm R}_{n}$,
so that we can estimate, for each QNM branch,
$\displaystyle N(\omega) \sim \frac{\omega}{\Delta \omega^{\rm R}}  = \frac{L}{c\pi}\omega$.
Multiplying the result by $2$ (for the two branches) we recover (\ref{e:Weyl_QNM_one-dim}).

The one-dimensional result (\ref{e:Weyl_QNM_one-dim}) is extended in \cite{Froes97}
(see also \cite{Simon00}) beyond the case of compact support potentials $V(x)$
to  a Weyl-like law valid for potentials
with a very fast decay, namely with a `super-exponentially decreasing' behaviour.
Specifically (cf.  expression in ``Conjecture 1.2'' in \cite{Froes97})
\bea
\label{e:Weyl_super-exponential}
N(\omega) \sim  \frac{s_V}{2\pi \rho} \omega^\rho \ \ , \ \ (\omega\to\infty) \ ,
\eea
where $s_V$ is a quantity determined by $V(x)$, in particular
its decay rate at infinity, and $\rho$ is the so-called `order $\rho$' of $V(x)$
(see e.g. \cite{Froes97}).
In particular, for functions referred to as of `exponential type' (see again \cite{Froes97}),
it holds $\rho=1$. Unfortunately, the theorem in \cite{Froes97} requires
the potential $V(x)$ to be super-exponentially decreasing, with $\rho>1$.
This shows in particular that for generic potentials, even those
satisfying a Weyl-like power law in $\omega$, the connection between
the power of $\omega$ and space dimension $d$ is generically lost.
This is illustrated by a class of potentials including 
Gaussians $V(x)=e^{-ax^2}$, for which Weyl's law (\ref{e:Weyl_super-exponential})
is shown to correspond to
\bea
N(\omega) \sim 2\frac{a}{c^2\pi } \omega^2 \ \ , \ \ (\omega\to\infty)
\eea
(we have introduced the factor $c^2$, as compared to the expression in \cite{Froes97}
to keep track of the physical dimensions). The situation
for more generic potentials is therefore quite open and depending on the
details of $V(x)$ and it does not get better
when increasing the space dimension $d$, as we will see below.

\subsubsection{Thermalisation time $T_{\mathrm{therm}}$ as `length scale'}
\label{s:therm_time}
As commented above, the  QNM Weyl's law (\ref{e:Weyl_QNM_one-dim})
presents a factor $2$ with respect to the standard Weyl's law (\ref{e:Weyl_law_onedim})
for bounded states.
This factor $2$ is naturally understood in terms of the existence of two QNM branches symmetric
with respect the imaginary axis in the $\omega$-complex plane: QNMs frequencies
$\omega_n = \omega_n^{\rm R} + i \omega^{\rm I}_n$ indeed
come in pairs, $\omega_n^\mp= \pm |\omega_n^{\rm R}| + i \omega^{\rm I}_n$, corresponding
to the two modes propagating respectively towards the left ($\omega_n^-$)  and
towards the right  ($\omega_n^+$). 

From a physical perspective, and in the context of a mode propagating to the
left and another propagating to the right, it is natural to interpret
the quantity $2 L/c$ as the time required for the thermalization of the system at the
scale $L$, namely the time to travel forth and back  the distance $L$
\bea
\label{e:thermalization_time}
T_{\mathrm{therm}} = 2\left(\frac{L}{c}\right) \ .
\eea
(see \cite{Liu:2021aqh}
for an interpretation along these lines). We will refer to 
 $T_{\mathrm{therm}}$ as the ``thermalization time''.
Such time scale, in contrast to a length scale, seems also the natural one
in a QNM setting employing a frequency
spectral parameter $\omega$, rather than a wavenumber $k$.
We can then rewrite (\ref{e:Weyl_QNM_one-dim}) as
\bea
\label{e:Weyl_QNM_one-dim_Time}
N(\omega) \sim  \left(\frac{T_{\mathrm{therm}}}{\pi}\right) \omega \ \ , \ \ (\omega\to\infty) \ .
\eea

\subsection{QNM Weyl's law: higher dimensions}
\label{s:Weyl_QNM_high-dim}
Considering QNMs of a compact support, bounded potential $V(x)$
in odd-dimensions $d\geq 3$, the following upper bound
for the QNM counting function holds~\cite{dyatlov2019mathematical}
\bea
\label{e:N_inequality}
N(\omega) \leq C_V \omega^d \ .
\eea
This provides a bound for $N(\omega)$ but, in contrast with the
one-dimensional case where the power in Weyl's law (\ref{e:Weyl_QNM_one-dim})
is indeed given by the dimension $d=1$, in the $d\geq 3$ case
this is not enough to establish Weyl's law in the large-$\omega$
asymptotics of $N(\omega)$ for generic compact support potentials.

An important exception to this is the radial case $V(x)=V(||x||)$.
Specifically, given a spherically-symmetric bounded, compact support
potential $V(||x||)$ in odd dimension $d\geq 3$, with
support inside a ball $B^d_R$ of radius $R$ (namely $\mathrm{chsupp}(V)=B^d_R$)
and with a jump at its boundary $V(R)\neq0$),
it holds the Weyl's law~\cite{Zwors89} 
\bea
\label{e:N_spherically_symmetric}
N(\omega) \sim C_R \omega^d + o(\omega^{d-1}) \ \ , \ \ (\omega\to\infty) \ .
\eea
This is a fine result. Remarkably, an (almost) explicit form can be given
for the constant $C_R$.
Specifically it holds~\cite{Stefa06}
\bea
\label{e:C_R}
C_R &=& \left(2  \frac{(\mathrm{Vol}_d(B^d_1))^2}{(2\pi)^d} + A_{S^{d-1}}\right) R^d  \nn \\
&=& \left(2  \frac{(\mathrm{Vol}_d(B^d_1))}{(2\pi)^d} \mathrm{Vol}_d(B^d_R) + A_{S^{d-1}} R^d\right) \nn \\
&=&  \left(2 C_d \mathrm{Vol}_d(\mathrm{chsupp}(V)) + A_{S^{d-1}} R^d\right)
 \eea
 where we have used the expression $\mathrm{Vol}_d(B^d_R)= \mathrm{Vol}_d(B^d_1)R^d$
 for the Euclidean volume of the ball of radius $R$. Here $A_{S^{d-1}}$
 is the constant in the QNM Weyl's law of the associated ``obstacle problem''.
 By the later we mean the $d$-dimensional scattering problem in the exterior of a $(d-1)$-sphere of radius $R$,
 namely $S_R^{d-1}$. That is, the QNM problem defined as
 \bea
\left\{
  \begin{array}{l}
    -\Delta \phi_n = \omega^2_n \phi_n  \ \ , \ \ \hbox{(in } \mathbb{R}^d\setminus B^d_R)\\
    \hbox{Homogeneous boundary conditions at } S_R^{d-1} \ , \\
    \hbox{Outgoing boundary conditions at infinity} \ ,
  \end{array}
\right.
    \eea
    for which a Weyl's law holds as~\cite{Stefa06}
    \bea
N(\omega) \sim A_{S^{d-1}} \omega^d + o(\omega^{d-1}) \ \ , \ \ (\omega\to\infty)
    \eea
    independently~\footnote{An explicit (but formal) expression can indeed be found for $A_{S^{d-1}}$.
    See Eq. (6) in \cite{Stefa06}.} of the Dirichlet, Neumann or Robin (homogenous) boundary conditions
    at $S_R^{d-1}$. Note the additive form of $C_R$ in expression (\ref{e:C_R}) with two terms
    \bea
    \label{e:C_R_splitting}
    C_R = C_R^V + C_R^{\mathrm{ext}} \ ,
    \eea
    where:
    \begin{itemize}
    \item[i)] $C_R^V= 2 C_d \mathrm{Vol}_d(\mathrm{chsupp}(V))$.
      This is a contribution associated with the potential $V(x)$.
      This is exactly twice the expression for the selfadjoint Weyl's law in a compact domain $D$,
      with the volume of $D$ substituted by the volume of the convex hull of the support of $V(x)$.
      
    \item[ii)]  $C_R^{\mathrm{ext}}= A_{S^{d-1}} R^d$. This is  contribution associated with the ``exterior problem'',
      in particular with $V(x)=0$.
        \end{itemize} 
  We will come back to this additive character of $C_R$, later in the discussion of the BH QNM Weyl's law.

  If we relax the compact support condition on the potential $V(x)$, we already lose control on
  the validity of the upper bound (\ref{e:N_inequality}) and a general Weyl's law is also out of
  reach, as dramatically illustrated in the one-dimensional case~\footnote{We have not
    discussed the even dimension $d$ case. References to bounds on $N(\omega)$ can be
    found in~\cite{Stefa06}.}. 

  The actual fact is that the problem of the  asymptotics of the QNM counting function $N(\omega)$
  for QNMs is much more involved than in the selfadjoint case.
  As stated in Ref.~\cite{dyatlov2019mathematical}: {\em ``Weyl laws for
counting resonant states are more complicated and richer [than for bound states] as they involve
both energy and rates of decay. Even the leading term can be affected by
dynamical properties of the system''}. Therefore in the most generic case, in particular
  for non-compact potentials in high-dimensions and with slow decay, a QNM Weyl's law sharing the
  universal features of the selfadjoint one is not expected. Again, citing~\cite{zworski2017mathematical}
  in reference to the QNM counting function:
  {\em ``Except in dimension one [...] the issue of asymptotics or even optimal lower bounds
remains unclear''.}

\section{QNM Weyl's law for black holes}
\label{s:QNM_Weyl_BHs}
As discussed at the end of section \ref{s:Weyl_QNM_high-dim}, the status of the
QNM Weyl's law in a general setting remains an open problem. This is in particular
the case for non-compactly supported potentials and for the discussion in
higher-dimensions. In this setting, it seems quite a remarkable feature
that QNM of $(3+1)$-dimensional BHs indeed satisfy a Weyl's
law~\cite{Jaramillo:2021tmt} analogous to the standard one 
of compact manifolds discussed in section \ref{s:Weyl_law}.

Stationary vacuum BHs are very particular solutions in Einstein theory,
as a consequence of the uniqueness theorem~\cite{Chrusciel2012}, showing
very special structural (integrability) features~\cite{Frolov2017}. One might ask then if
satisfying the standard Weyl's law, with a power-law controlled
by the space dimension $d$, is a consequence of such special character.
However, such concerns seem not to apply for the small scale
(high-wavenumber) features of BH spacetimes, that are the relevant ones in the high-frequency asymptotics
of the QNM counting function $N(\omega)$. In addition, as discussed in~\cite{Jaramillo:2021tmt}, the
Weyl's (power-)law is robust under high-frequency (actually high-wavenumber) perturbations,
in spite of the ultraviolet instability of QNM
overtones~\cite{alsheikh:tel-04116011,Jaramillo:2020tuu,Jaramillo:2021tmt,Gasperin:2021kfv,Jaramillo:2022kuv,Destounis:2021lum, boyanov2023pseudospectrum,sarkar2023perturbing,Arean:2023ejh,Destounis:2023ruj}.
This strongly suggests that the Weyl's law findings in~\cite{Jaramillo:2021tmt} are indeed 
valid for generic BH spacetimes, in particular independent of the asymptotics
at infinity.

In this section we detail and extend the discussion of the
BH QNM Weyl's law in \cite{Jaramillo:2021tmt}, adopting the heuristic
methodology there based on the explicit evaluation of QNMs,
as eigenvalues of a non-selfadjoint operator. For concreteness,
the calculation is done in the $(3+1)$-spherically symmetric asymptotically flat
case, though the results indeed extend to more generic settings.
Very importantly, no proof is provided of the presented results, but rather 
a straightforward  verification of the standard Weyl's law (\ref{e:BH_QNM_Weyl})
taken as an Ansatz for the asymptotics of the BH QNM $N(\omega)$.
As commented in section \ref{s:Invitation_BH_QNM_Weyl}, this is in stark
contrast with the rigorous  results in~\cite{Dyatlov:2013hba,Dyatlov:2013fua}
on a Weyl's law for Kerr-de Sitter BHs. In section \ref{s:Dyatlov-Zworski}
we briefly sketch these semi-classical results by Dyatlov \& Zworski
and comment on the relation with the results here presented.

\subsection{Effective one-dimensional BH potentials} 
\label{s:QNM_Weyl_one-dim}
We focus our Weyl's law discussion in the $(3+1)$-Schwarzschild BH, proceeding
in several stages from the reduction to a one-dimensional effective problem
to the full three dimensional Weyl's law. Specifically, we consider
here QNM frequencies associated to gravitational perturbations.

Making use of its spherical symmetry, gravitational perturbations in
Schwarzschild can be reduced to two independent scalar one-dimensional
wave equations of the form (\ref{e:QNM_wave}), with $x$ the so-called
tortoise coordinate $r_*\in]-\infty, \infty[$
    \bea
    \label{e:QNM_Schwarz}
    \left(-\frac{d^2}{dr_*^2} + V(r(r_*))\right) \phi_{n\ell m} = \omega^2_{n\ell m}\phi_{n\ell m}  \ ,
    \eea
where the angular modes $(\ell, m)$ correspond to the spherical mode decomposition.  
    More explicitly, the 
    ``axial'' and ``polar'' Schwarzschild gravitational parities
    have respectively associated the  Regger-Wheeler
$V^{\mathrm{RW},s}_{\ell}(r)$ and
    Zerilli $V^{\mathrm{Z}}_{\ell}(r)$ potentials
    \cite{Regge57,Zeril70,Chandrasekhar:579245,Kokkotas:1999bd,maggiore2018gravitational},
    with $r$ the radial Schwarzschild coordinate 
    \bea
\label{e:Schwarzschild_potential_RG}
V^{\mathrm{RW},s}_{\ell}(r) =  \left(1-\frac{2M}{r}\right)  \left(\frac{\ell(\ell + 1)}{r^2} +(1-s^2) \frac{2M}{r^3} \right)\ , \nonumber\\
\eea
for the axial case ($s=0,1,2$ correspond to the scalar, electromagnetic
and (linearized) gravitational cases), and
\bea
\label{e:Schwarzschild_potential_Z}
&&V^{\mathrm{Z}}_{\ell}(r) =  \left(1-\frac{2M}{r}\right) \nn \\
&&\left(\frac{2n^2(n+1)r^3 + 6n^2Mr^2+18nM^2r+18M^3}{r^3(nr+3M)^2} \right) \ , \nonumber\\
\eea
with
\bea
n = \dfrac{(\ell-1)(\ell+2)}{2} \ ,
\eea
for the polar case (the expression $r_* = r + 2M\ln(r/2M-1)$ gives the
relation between the $r_*$ and the standard radial Schwarzschild coordinate $r$
in the potentials' expressions).

A key feature of Schwarzschild is its so-called QNM isospectrality, namely,
QNM frequencies associated with the $V^{\mathrm{RW},s}_{\ell}(r)$ and $V^{\mathrm{Z}}_{\ell}(r)$
coincide (see \cite{Lenzi:2021njy,Lenzi:2021wpc,Lenzi:2022wjv,Lenzi:2023inn} for a fine
discussion of this issue in terms of underlying integrability properties;
see also references therein). We can therefore choose
either of the two potentials above in our discussion. For simplicity one can
think of $V^{\mathrm{RW},s}$ with $s=2$ (gravitational case).

The key point in our discussion is that the relevant one-dimensional potential
is of non-compact support.
Therefore it is not in the class of potentials under the hypothesis of the
theorem in \cite{Zwors87} leading to the Weyl's asymptotics (\ref{e:Weyl_QNM_one-dim}).
Even more, when expressed in terms of the $r_*$ coordinate entering in
the QNM equation (\ref{e:QNM_Schwarz}), the decay is only exponential at $r_*\to -\infty$
and, even worse, a power-law at $r_*\to \infty$. Such decays are therefore
too slow and do not fall in the hypotheses of the theorem in \cite{Froes97}
leading to the asymptotics (\ref{e:Weyl_super-exponential}). Therefore, there is no reason
a priori to expect that the asymptotics of the QNM counting function $N(\omega)$
for the Schwarzschild BH satisfies a Weyl's law. And, in spite of this, it does.

To show it, let us consider the large-$n$ QNM asymptotics of $\omega_{n\ell m}$
for a fixed mode $(\ell m)$ (QNM frequencies are actually degenerated in
the azimuthal mode $m$, but we keep it in the present QNM counting context).
Following Nollert~\cite{Nollert:1993zz} 
\bea
\label{e:asymptotics_n_Schwarz_QNM}
\!\!\!\! 2M\omega_{n\ell m} &=&  \pm 0.0874247 + i\frac{1}{2}\left(n - \frac{1}{2}\right) + \ldots  \\
&=& \pm \frac{\ln(3)}{4\pi} + i\frac{1}{2}\left(n - \frac{1}{2}\right) + \ldots \ \ (n \gg 1) \ ,
\eea
that it is not only independent of $m$, but also of $\ell$.

\begin{figure}[t!]
\centering
\includegraphics[width=8.4cm]{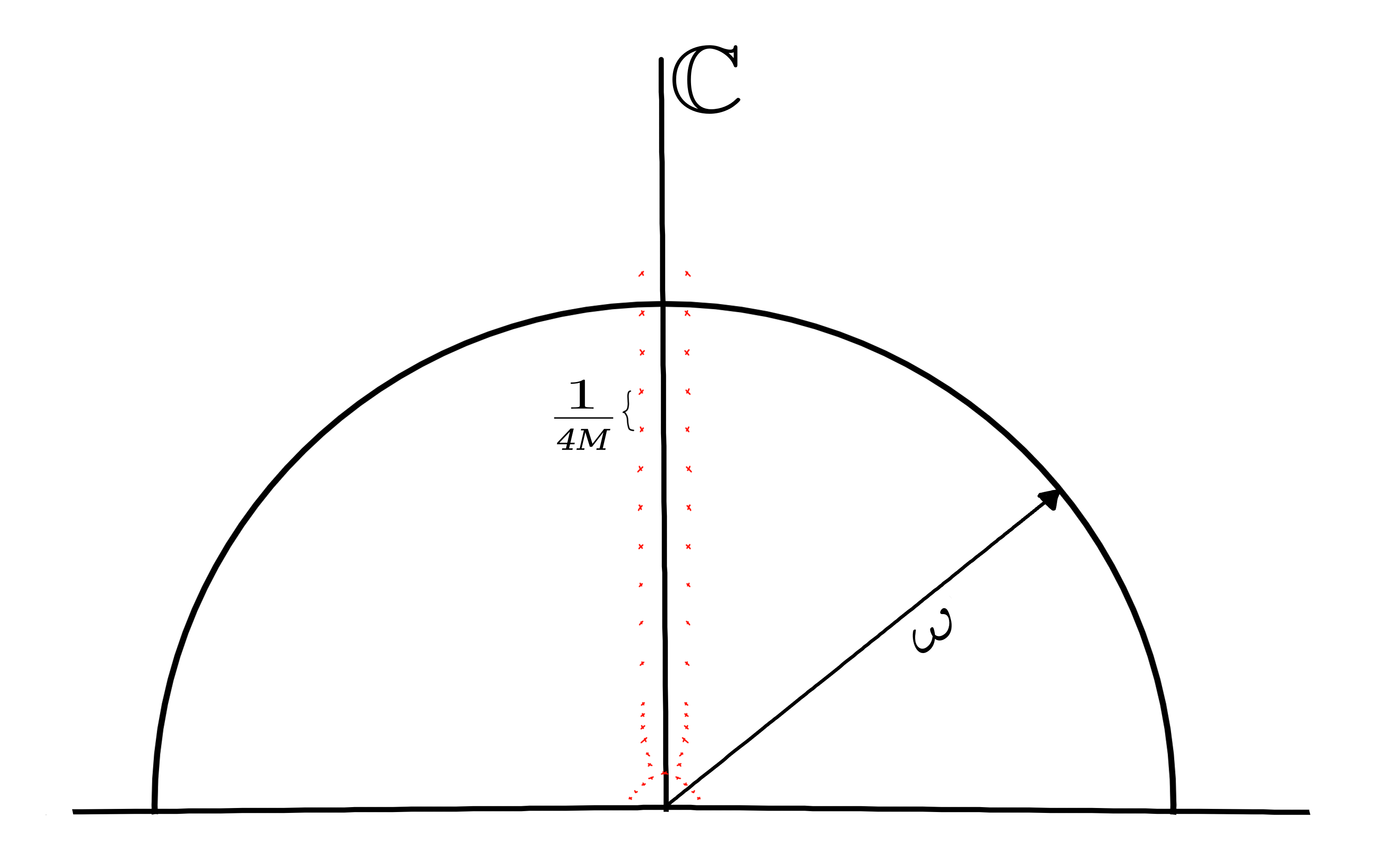}
\caption{
  Sketch of the figure used in the ``counting argument''
  of QNMs in the Schwarzschild BHs. The key features are
  the asymptotic distribution  with a constant gap of QNMs along a line parallel
  to the imaginary axis, as well as the
  bound on the real part of QNM frequencies $\omega_n$'s
  in terms of the real part of the fundamental QNM.
 }
\label{fig:QNM_Schw_l-fixed}
\end{figure}

As sketched in Fig.~\ref{fig:QNM_Schw_l-fixed}, 
 asymptotically
QNMs are placed  parallel to the imaginary axis.
Therefore, for sufficiently large $n$, the asymptotics of $N_{\ell m}$  are  
controlled by (\ref{e:asymptotics_n_Schwarz_QNM}) in a simple manner.
Concretely, the  QNM frequencies $\omega_{n\ell m}$ with
fixed $(\ell m)$ are homogeneously distributed
parallel to the imaginary axis, with a constant vertical gap
\bea
\label{e:Delta_omega}
\Delta  \omega_{n\ell m} =\omega_{(n+1)\ell m}-\omega_{n \ell m}=\frac{1}{4M} \ ,
\eea
given by the BH surface gravity, $\kappa=1/(4M)$ for the Schwarzschild
BH.
Therefore, for each $(\ell, m)$ and each one of the two QNM branches
parallel to the imaginary axis, the number $N_{\ell m}(\omega)$
of QNMs contained in that circle of radius $\omega$ follows from (\ref{e:asymptotics_n_Schwarz_QNM})
and is approximately given by 
\bea
N_{\ell m}(\omega) \sim \frac{2M \omega}{1/2} = 4 M \omega  \ \ , \ \ (\omega\to\infty) \ .
\eea
Counting the two branches, we finally have
\bea
\label{e:Weyl_Schwarz_one-dim}
N_{\ell m}(\omega) \sim 8M \omega \ \ , \ \ (\omega\to\infty) \ ,
\eea
that can be rewritten in term of the surface gravity as
\bea
\label{e:Weyl_Schwarz_one-dim_kappa}
N_{\ell m}(\omega) \sim 2\cdot \frac{1}{\kappa}\cdot \omega \ \ , \ \ (\omega\to\infty) \ ,
\eea
This satisfies indeed the one-dimensional Weyl's law (\ref{e:Weyl_QNM_one-dim}),
with the proper $d=1$ dimensionality in the power law and a
length scale independent of $(\ell m)$ angular numbers,
fixed  (putting $c=1$) by the relation $2 L/\pi = 8 M = 2/\kappa$, that is
\bea
\label{e:L_one-dim}
L=4\pi M = \frac{\pi}{\kappa}\ \ .
\eea

We do not have a first-principles justification for this one-dimensional BH QNM Weyl's law,
namely expressions (\ref{e:Weyl_Schwarz_one-dim}) or (\ref{e:Weyl_Schwarz_one-dim_kappa}),
apart from the presented direct
verification of large $n$ QNM asymptotics (\ref{e:asymptotics_n_Schwarz_QNM}) found by Nollert.
In particular, the length $L$ in (\ref{e:L_one-dim}) stands as  an effective one fixed by the
potential.
Some physical speculative heuristics about $L$ 
are presented in~\cite{Jaramillo:2021tmt}. 
But more interestingly and remarkably, such an effective length determined by the potential is
characterised by a purely  geometric BH horizon property, the surface gravity $\kappa$.
This follows from the role of the latter in the asymptotic gap (\ref{e:Delta_omega}) in the imaginary
part $\mathrm{Im}(\omega_n)$ of BH QNM frequencies $\omega_n$'s. Such gap $\Delta \mathrm{Im}(\omega)\sim\kappa$
is a robust feature of this discussion and plays a fundamental role in the very characterisation
of BH QNMs, as shown in \cite{Warnick:2013hba} in de Anti-de Sitter context. We will encounter
$\kappa$ again below in our discussion.

\subsubsection{Weyl's law robustness under ultraviolet BH QNM instability}
\label{s:Weyl_perturbed_BHs}
As commented above and presented in~\cite{Jaramillo:2021tmt}, the
Weyl's law (\ref{e:Weyl_Schwarz_one-dim}) is robust under `ultraviolet'
perturbations of the effective potentials (\ref{e:Schwarzschild_potential_RG})
and (\ref{e:Schwarzschild_potential_Z}). More specifically, under 
high-wavenumber (small scales) tiny perturbations of $V(r)$, the large-$n$
asymptotics of $N(\omega$) keeps a power law with the same power $d=1$,
though the multiplicative constant depends on the perturbation
scale (fixed by its wave number), transitioning from the Schwarzschild
value to approximately three times that value (cf. behaviour of $L^{\mathrm{W}}$
in Fig. 2 of \cite{Jaramillo:2021tmt}).

It is a remarkable feature that such robustness of the Weyl's law
under ultraviolet perturbations happens in spite of the completely
different distribution in the complex plane of unperturbed and perturbed QNM overtones:
whereas in the unperturbed case QNM overtones distribute asymptotically
in parallel to the imaginary axis, in the perturbed case they migrate and 'open'
to new branches in the
complex plane, with unbounded real part, distributing
themselves along (Nollert-Price-like) asymptotically open logarithmic
branches~\footnote{Actually, together with this logarithmic QNM branches,
  another family of ``inner QNMs'' appear. The latter seem to play a key
  role in the change of the multiplicative length scale in the Weyl's law~\cite{Jaramillo:2021tmt}.
  This will be systematically studied elsewhere.}.

The rationale for the preservation of Weyl's law in the perturbed case
seems to be intimately related to the structure
of Regge QNMs in (\ref{e:Regge_branches}) that, as discussed above,
encodes the Weyl's law in dimension one. Specifically, Weyl's law in the
ultraviolet perturbed case would be a consequence of the conjecture
proposed in \cite{Jaramillo:2021tmt} and refined in 
\cite{Gasperin:2021kfv}, according to which the ultraviolet QNM overtone instability
is a general relativity low-regularity phenomenon in which perturbed
Nollert-Price overtone branches tend towards Regge QNM branches
in the infinite wavenumber limit of the perturbations.
The key point is that, although the branches dramatically migrate
to different regions in the complex plane, the distribution of
QNM overtones along the branches remains homogeneous with an
approximately constant separation along the branch, so that the
simple counting argument presented above and illustrated in Fig.~\ref{fig:QNM_Schw_l-fixed}
remains essentially unchanged. The analysis of this limit
and its consequences on Weyl's law, in particular the dependence of the
length scale in the Weyl's law with the ``small scale'' of the perturbation, will
be presented elsewhere~\cite{Besson_et_al}.

\subsection{Three dimensional case: analytic estimation}
\label{s:3-dim_case_analytical}
We assess now the Weyl's law in the full
$(3+1)$-dimensional Schwarzschild BH, in particular
recovering the power $d=3$. The argument is not fine
enough to correctly estimate the multiplicative volume
scale in Weyl's law, see below in section \ref{s:3-dim_case_numerical},
but it offers a non-trivial insight.

Expression (\ref{e:Weyl_Schwarz_one-dim}) provides the QNM counting
function for a fixed $(\ell, m)$ mode. Let us consider now the sum 
over $(\ell, m)$ 
\bea
\label{e:QNM_counting_1}
N(\omega) &=& \sum_{\ell=2}^{\ell_{\mathrm{max}}} \sum_{m=-\ell}^{\ell}N_{\ell m}
\sim \sum_{\ell=2}^{\ell_{\mathrm{max}}}  \sum_{m=-\ell}^{\ell} 8 M \omega \nn \\
&=& 8 M \omega \sum_{\ell=2}^{\ell_{\mathrm{max}}} (2\ell + 1) \ ,
\eea
where we have made use of the degeneracy in the azimuthal number $m$.
The key point is to estimate $\ell_{\mathrm{max}}$.
This depends critically on the specific
distribution of QNMs in the complex plane for Schwarzschild.
Such structure is illustrated, for instance, in the upper panel
of Fig.~\ref{fig:4DBH_WeylLaw}:
the real part of the two QNM branches of fixed $\ell$ is bounded by their value
for the ``branch fundamental mode'' $\omega_{0,\ell}$, then they
cross at the algebraically special QNM on the imaginary axis and 
asymptote to a line parallel to the imaginary axis. The key point
is that for a given circle of radius $\omega$, all QNM branches
with $\mathrm{Re}(\omega_{0,\ell})>\omega$ lie outside that
circle (note that horizontal and vertical scales in the figure
are the same, so the depicted circle is faithful to the
present argument), so they do not contribute to $N(\omega)$.
For a given $\ell$, we can estimate the value of
$\mathrm{Re}(\omega_{0,\ell})$ from the large-$\ell$ asymptotics
(cf. e.g. Eq. (32) in~\cite{Kokkotas:1999bd})
\bea
3\sqrt{3}M\omega_{n,\ell} \sim \ell + \frac{1}{2} + i\left(n +\frac{1}{2}\right) \ \ , \ \ (\ell\to\infty)
\eea
that is intimately related to the light ring radius, $r=3M$.
Therefore, we can write for the real part (for $n=0$)
\bea
3\sqrt{3}M \mathrm{Re}(\omega_{0,\ell}) \sim \ell + \frac{1}{2} \ \ , \ \ (\ell\to\infty) \ .
\eea
As argued above, branches with $\mathrm{Re}(\omega_{0,\ell})>\omega$ do not
enter in the QNM counting, so QNMs with $\ell> 3\sqrt{3}M \omega$ do not
contribute to $N(\omega)$, this providing the estimate
\bea
\label{e:l_max}
\ell_{\mathrm{max}} \sim 3\sqrt{3}M \omega \ .
\eea
As seen in Fig.~\ref{fig:4DBH_WeylLaw}, this overestimates the counting of QNMs
for large $\ell$ close to $\ell_{\mathrm{max}}$, but has the virtue of relating the
counting to the light ring structure. We can then write (\ref{e:QNM_counting_1}) as
\bea
N(\omega) \sim 8M \omega \sum_{\ell=2}^{3\sqrt{3}M\omega} (2\ell + 1)
 \sim  8M \omega \cdot 2 \sum_{\ell=2}^{3\sqrt{3}M\omega} \ell \ .
 \eea
 Using $\displaystyle \sum_{i=2}^p i \sim \sum_{i=1}^p i = \frac{p(p+1)}{2}\sim \frac{p^2}{2}$
 (for $p\gg 1$), we get
 \bea
 \label{e:N_analytical_estimation}
 N(\omega) \sim 8M \omega \cdot 2 \frac{27M^2 \omega^2}{2} = (6M)^3 \omega^3
 \eea
 As commented above, the factor $(6M)^3$ overestimates $N(\omega)$, so we should
 rather keep as a proper result of this argument
 \bea
 N \sim C_{\mathrm{Schwarz}}\;\omega^3 \ ,
 \eea
 with  $C_{\mathrm{Schwarz}}$ a constant. We proceed now
 to estimate it from the direct counting in numerically calculated QMNs.

\subsection{Three dimensional case: numerical computation}
\label{s:3-dim_case_numerical}
\begin{figure}[t!]
\centering
\includegraphics[width=8.4cm]{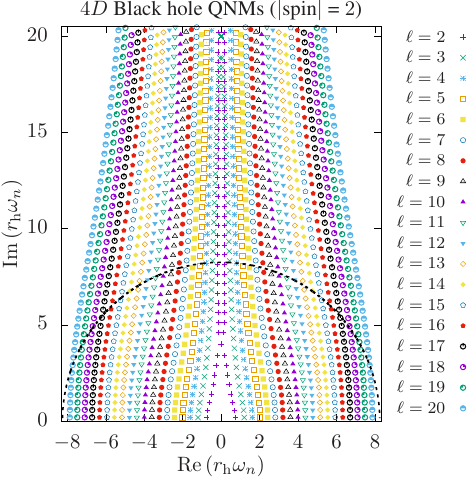}
\\[3ex]
\includegraphics[width=8.4cm]{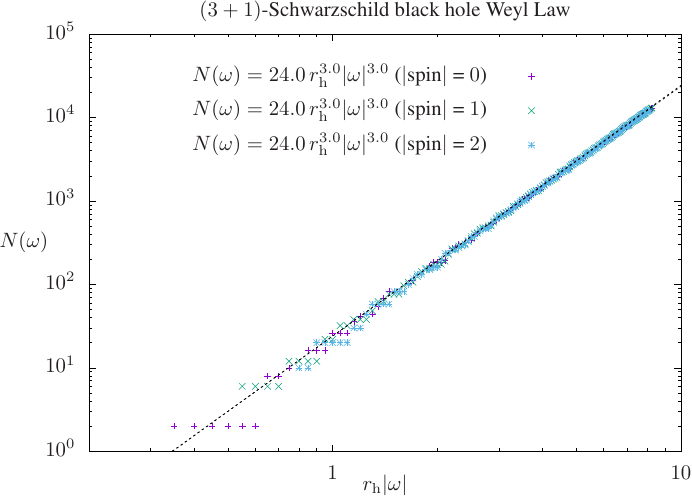}

\caption{ Numerical study of the QNM Weyl's law for Schwarzschild. 
  {\em Top panel:} Numerically calculated BH QNMs for a gravitational perturbation ($s$=$|$spin$|$ = 2) on the $(3+1)$-dimensional
  Schwarzschild spacetime. Within the circle $|r_{\rm h} \omega| = 8.1$, with $r_{\rm h}=2M$,
  one counts the number of QNMs for the angular modes $\ell=2,\cdots, 20$. {\em Bottom panel:} Weyl's law for $N(\omega)$ in $(3+1)$-dimensional
  Schwarzschild BHs for scalar ($s$=$|$spin$|$ = 0), electromagnetic ($s$=$|$spin$|$ = 1) and gravitational ($s$=$|$spin$|$ = 2) perturbations.
  The asymptotic $N(\omega) \sim |\omega|^3$ is recovered, with a multiplicative constant
  $\approx 24.0 \; r_{\rm h}^{3.0}$.
 }
\label{fig:4DBH_WeylLaw}
\end{figure}

Using the hyperboloidal, compactified scheme in
\cite{Ansorg:2016ztf,PanossoMacedo:2018hab,PanossoMacedo:2018gvw,Jaramillo:2020tuu,Jaramillo:2021tmt}
for the numerical calculation
of QNMs in a spherically symmetric setting, the result $(\ref{e:N_numerical_estimation})$ follows from
a straightforward computation (with $2\leq\ell \leq 20$ in the gravitational
case, $s=2$, cf. top panel in Fig.~\ref{fig:4DBH_WeylLaw})
\bea
\label{e:N_numerical_estimation}
N(s) = \alpha \cdot (4M)^3 \omega^{3.0} \ ,
\eea
with $\alpha\sim 2.98$. On the one hand, this is consistent with the power
law $d=3$, that we recovered with the analytic estimation in
(\ref{e:N_analytical_estimation}). Regarding the multiplicative factor, 
it is tempting to approach $\alpha\sim 3$.
Confirming this requires the use of more $\ell$'s in the calculation and
this becomes rapidly challenging from a numerical point of view.
Under such assumption $\alpha\sim 3$, the analytical estimation (\ref{e:N_analytical_estimation})
for the counting function,  that we will refer as $N^\mathrm{analytic}(\omega)$, overestimates
the numerical estimation $N^\mathrm{numerical}(\omega)$  in (\ref{e:N_numerical_estimation}) 
by 
\bea
\label{e:Buchdahl}
\frac{N^\mathrm{analytic}(\omega)}{N^\mathrm{numerical}(\omega)}
\sim \frac{(6M)^3 \omega^3}{3(4M)^3 \omega^3} = \frac{9}{8} \ .
\eea
Given the approximations in the argument leading to (\ref{e:N_analytical_estimation}),
this seems quite a reasonable estimation, thus supporting the role
of the light-ring scale, $L^{\mathrm{LR}} = 3M$, as a relevant one
in the BH QNM counting problem~\footnote{As a curious remark, the factor $9/8$ is the one
  appearing in the so-called Buchdahl bound requiring
$R > (9/8)R_{\mathrm{Schwarz}}$ ($R$ is the areal radius) for
the stability of static, spherically symmetric matter configurations.
If this Buchdahl bound plays a role in our QNM discussion, it is for now intriguing but unclear.}.
This is consistent with the
key role played by the light ring for BH QNMs (cf. e.g.~\cite{Berti:2009kk}). We will
come back to this light trapping point in section \ref{s:Dyatlov-Zworski}.

Finally, the bottom panel of Fig.~\ref{fig:4DBH_WeylLaw} shows the
results of the Weyl's law for all types of perturbations in
the Regge-Wheeler potential (\ref{e:Schwarzschild_potential_RG}),
namely  scalar ($s=0$), electromagnetic ($s=1$) and gravitational ($s = 2$)
perturbations. The obtained Weyl's law seem to be independent
of the type of perturbation, in spite of the differences in the
associated potentials, supporting the idea that the
multiplicative scale is a geometric property intrinsic
to the underlying Schwarzschild BH spacetime.

\subsection{BH QNM Weyl's law: redshift and trapping}
\label{s:BH_QNM_proposal}
\begin{figure}[h!]
\centering
\includegraphics[width=8.4cm]{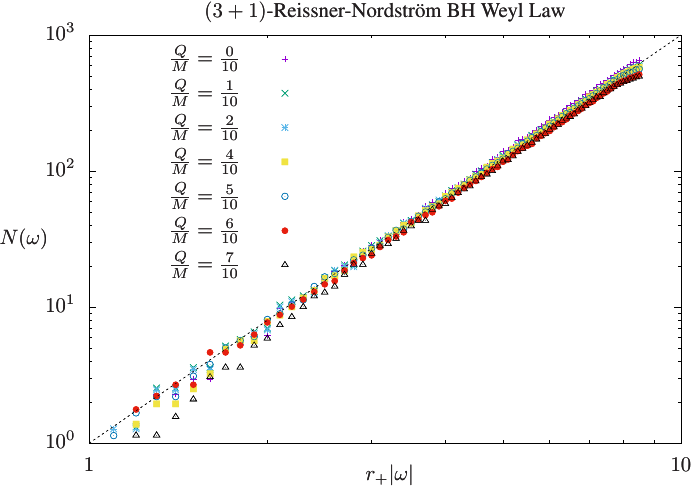}

\caption{Numerical study of the QNM Weyl's law for the  $(3+1)$-Reissner-Nordstr\"om BH.
  The Weyl's law asymptotics for $N(\omega)$ for QNMs
  of gravitational ($s$=$|$spin$|$ = 2) perturbations are presented for different values
   of $Q/M$ (charge/mass ratio). We recover the asymptotic $N(\omega) \sim |\omega|^3$. However,
  contamination from the branch cut eigenvalues in our numerical scheme prevents us from recovering
  the multiplicative constant, that we have normalised to $1$ for all $Q/M$'s.
 }
\label{fig:4DBH_WeylLaw_RN}
\end{figure}

We have just seen, both using an analytic estimation or a straightforward
numerical calculation, that the asymptotics of the counting function $N(\omega)$
for the Schwarzschild BH is consistent with the classical Weyl's law,
with a power law given in terms of the dimension $d$ of spatial sections
of the spacetime.
Further support  is provided by the Reissner-Nordstr\"om case
  for all values of the charge/mass ratio $Q/M$, as
  presented in Fig.~\ref{fig:4DBH_WeylLaw_RN}, where the exponent of the power-law is also
  numerically verified to be $d\approx 3$~\footnote{The determination of the multiplicative
    constant in our numerical scheme is however more delicate, due to the contamination
    of QNMs by spurious eigenvalues in the branch cut as the ratio $Q/M$ increases approaching
    extremality $Q/M\to 1$.    
    This issue can be by-passed by working with de Sitter or Anti-de Sitter asymptotics
    and will be presented elsewhere~\cite{Besson_et_al}.}.
 Given the spherical symmetry of the Schwarzschild and Reissner-Nordstr\"om BHs, this
result is certainly suggested by the asymptotics for spherically symmetric potentials
in odd dimensions $d$ discussed in section~\ref{s:Weyl_QNM_high-dim},
namely expression~(\ref{e:N_spherically_symmetric}) for $N(\omega)$.
However there is a key difference~\footnote{Another minor point to
  highlight is that the Schwarzschild QNM problem is not of the
  form $(-\Delta + V(r))\phi_n = \omega^2\phi_n$ for the Euclidean
  Laplacian $-\Delta$ in references~\cite{Zwors89,Stefa06}.}: in contrast with the (key)
assumption in~\cite{Zwors89,Stefa06} on the compact support of the potential,
the potentials  $V(r)$ in (\ref{e:Schwarzschild_potential_RG})
and (\ref{e:Schwarzschild_potential_Z}) (as well as the corresponding potential for the Reissner-Nordstr\"om BH)
are not only of non-compact
  support, but they present a slow power-law decay. It is therefore remarkable that
  the standard selfadjoint Weyl's law holds. We conjecture that
  this is indeed the  case for generic (not only asymptotically flat and not only spherically
  symmetric) BH $(d+1)$-dimensional
  spacetimes,  with odd $d$, so the asymptotics (\ref{e:BH_QNM_Weyl}) holds for BH QNMs.

  This is a bold statement, given the special nature of the spherically symmetric
  case and in the absence of a study specifically devoted to a non-spherically
  symmetric case, namely Kerr. But it is also in the spirit of the conjecture
  of Weyl's law by Sommerfeld~\cite{Somme10} and Lorentz~\cite{Loren10}
  preceding (and leading to) the Weyl's proof one year later \cite{Weyl11,Weyl12} (cf. e.g.
  the discussion in \cite{AreNitPet09}). In the specific case of BH spacetimes,
  such a proposal is encouraged by the genericity and universality properties
  of stationary BH  vacuum solutions. Such a universality is akin to the
  universality of Weyl's law.
  
  Essentially motivated by structural consistency with the standard Weyl's law
  (\ref{e:counting_function_compact}) for bound states of selfadjoint operators
  in compact manifolds, we can refine a bit further our proposal (\ref{e:BH_QNM_Weyl})
  in what refers to the volume factor. Specifically, we start from (\ref{e:N_spherically_symmetric})
  and (\ref{e:C_R}) and consider the additive splitting of $C_R$ in (\ref{e:C_R_splitting}).
  As discussed after Eq. (\ref{e:C_R_splitting}), there is a contribution $C_R^V$
  from the (convex hull of the) compact-support potential and a contribution $C_R^{\mathrm{ext}}$
  from the ``exterior'' scattering problem without potential (an ``obstacle'').
  We can consider now pushing further and further away the boundary of the  support of $V(||x||)$.
  Then, in a formal sense, non-compact support potentials  (as it is the case for
  Schwarzschild) can be seen as the (singular) limit in which the support of $V(||x||)$ is pushed
  to infinity and no ``exterior region'' remains. In particular, no contribution of $C_R^{\mathrm{ext}}$
  would enter into $C_R$, the latter being then given solely by $C_R^V= 2 C_d \mathrm{Vol}_d(\mathrm{chsupp}(V))$.
  Of course, this is just a formal expression, since $\mathrm{Vol}_d(\mathrm{chsupp}(V))$
  diverges and it should be substituted by an effective volume term $\mathrm{Vol}_d^{\mathrm{eff}}$
  completely fixed by the BH geometry.
  
  Upon these considerations, our proposal for the QNM Weyl law for $(d+1)$-dimensional
  BH spacetimes, with $d$ odd, is
\bea
\label{e:BH_QNM_Weyl_v2}
N(\omega) \sim \frac{2}{c^d}C_d\mathrm{Vol}_d^{\mathrm{eff}} \omega^d +  o(\omega^{d-1}) \ \ , \ \ (\omega\to \infty) \ .
\eea
We stress that $\mathrm{Vol}_d^{\mathrm{eff}}$ is a spacetime quantity, not linked with
a particular spatial foliation of the spacetime: it should be related to the
intrinsic properties of the spacetime geometry.

\subsubsection{Schwarzschild QNM Weyl's law: `thermal' heuristics}
\label{s:QNM_Schwarz_revisited}
We can revisit now the discussion in section \ref{s:3-dim_case_numerical} under the light of
the proposal (\ref{e:BH_QNM_Weyl_v2}).
We can determine $\mathrm{Vol}_d^{\mathrm{eff}}$ for the Schwarzschild BH in (\ref{e:BH_QNM_Weyl_v2}),
from the expression for $N(\omega)$ in (\ref{e:N_numerical_estimation})
and from the value of $C_d$ for $d=3$, namely $\displaystyle C_3 = \frac{1}{6\pi^2}$
(cf. (\ref{e:C_3})).
Adopting $\alpha=3$ in (\ref{e:N_numerical_estimation}), we have the Weyl's law
$N(\omega)\sim 3\cdot(4 M)^3\omega^3 + o(\omega^2)$. From
this it follows
\bea
\label{e:Vol_eff_Schwarz}
\mathrm{Vol}_3^{\mathrm{eff}} = (3\pi)^2 (4M)^3 \ .
\eea
This expression  is not very illuminating. We consider an
alternative point of view by making use of the thermalization time $T_{\mathrm{therm}}$
introduced in (\ref{e:thermalization_time}). From a physical perspective
(and in the spirit of the reasoning in the black body problem~\cite{AreNitPet09}) ,
instead of thinking in terms of a density of states $\rho(\omega)$
\bea
\label{e:Density}
\rho(\omega) \sim \frac{N(\omega)}{\mathrm{Volume}} \ ,
\eea
we can consider the related notion of radiation flux per unit of time and area, that
we denote by $F(\omega)$, namely
\bea
\label{e:Flux}
F(\omega) \sim  \frac{N(\omega)}{\mathrm{Time}\cdot \mathrm{Area}} \ .
\eea
Under this perspective, the multiplicative factor in the Weyl's
law has naturally the dimensions $[\mathrm{Time}]\cdot [\mathrm{Area}]$.
We can rewrite the factor in (\ref{e:BH_QNM_Weyl_v2}), for $d=3$,
in terms of the product of the thermalization time $T_{\mathrm{therm}}$ and  an effective
area $\mathrm{Area}^{\mathrm{eff}}$
\bea
\label{e:fixing_T}
\frac{2}{c^3}C_3\mathrm{Vol}_3^{\mathrm{eff}} = \frac{1}{c^{2}}C_3 T_{\mathrm{therm}} \mathrm{Area}^{\mathrm{eff}} \ .
\eea
In our spherically symmetric setting, this 
poses question about the ``natural sphere'' for evaluating $\mathrm{Area}^{\mathrm{eff}}$.
Invoking the role of the light-ring radius $R = 3M$ discussed in sections
\ref{s:3-dim_case_analytical} and \ref{s:3-dim_case_numerical} as a
relevant (key) scale in the Schwarzschild QNM counting problem (specifically, cf. discussion
after Eqs. (\ref{e:l_max}) and (\ref{e:Buchdahl})), we adopt a
``light ring'' prescription for $\mathrm{Area}^{\mathrm{eff}}$ 
\bea
\label{e:area-effr}
\mathrm{Area}^{\mathrm{eff}} = \mathrm{Area}^{\mathrm{LR}} = 4\pi\cdot (3M)^2 \ .
\eea
This expression for $\mathrm{Area}^{\mathrm{eff}}$, together with Eq. (\ref{e:fixing_T}) fixes the value
of the thermalization time, resulting in
\bea
\label{e:t_term_32pi}
T_{\mathrm{therm}} = 32\pi M \ .
\eea
This is a suggestive result, both for its simplicity and its natural
connection to BH `thermal aspects'. Indeed, regarding the ``radial scale''
in $\mathrm{Vol}_3^{\mathrm{eff}}$ as a ``length'', rather than as a ``time'', it is not natural
to explain the presence of the ``factor `$\pi$'', that appears
naturally related to ``angular lengths''. On the contrary when
looking at it as a ``time'', and moreover as a `thermalization time'
(the latter crucially related to the factor $2$ in QNM Weyl's laws), the
factor $\pi$ appears naturally as connected to the thermal
aspects of BH physics. 
Indeed, the factor $\displaystyle\frac{\kappa}{8\pi}$
 in BH first law (remind, $\kappa=(4M)^{-1}$ the BH surface gravity)
 \bea
 \label{e:BH_1st}
\delta M = \left(\frac{\kappa}{8\pi}\right)\cdot \delta \mathrm{Area}^{\mathrm{Hor}} \ ,
\eea
is precisely the inverse of the thermalization time $T_{\mathrm{therm}}$
(\ref{e:t_term_32pi})
\bea
\label{e:T_therm_kappa}
T_{\mathrm{therm}} = \frac{8\pi}{\kappa} \ .
\eea
In this context, at the interphase between BH QNMs and thermodynamical questions,
we note the relation between the thermalization $T_{\mathrm{therm}}$ in Eq. (\ref{e:T_therm_kappa})
and the BH relaxation time $\tau_{\mathrm{relax}}$ (given by the inverse of the imaginary part of
the fundamental (least damped) BH QNM frequency), in the setting of
Hod's `time-times-temperature' (TTT) conjecture~\cite{hod2007universal,hod2021quasinormal}
\bea
\tau_{\mathrm{relax}} \cdot T_{\mathrm{BH}} \geq \frac{\hbar}{\pi} \ ,
\eea
where  $T_{\mathrm{BH}}$ is Hawking BH temperature, given by
\bea
\label{e:T_Hawking}
T_{\mathrm{BH}} = \frac{\kappa\hbar}{2\pi} \ .
\eea
From this expression and Eqs. (\ref{e:T_therm_kappa}) and (\ref{e:T_Hawking}) it
follows~\footnote{One could actually argue that the factor $\pi$ suggests a link
  of $T_{\mathrm{therm}}$ with a temperature, rather than with a time,
  by using arguments of periodicity of the BH solutions in Euclidean time
relating time and temperature.
 We note that heuristic speculations on the possible connections between
$T_{\mathrm{therm}}$  and the Hawking temperature $T_{\mathrm{Hawking}}$ in the the BH evaporation
process were suggested in~\cite{Jaramillo:2021tmt}. Here we prefer to remain `sober' in our
discussion, leaving heuristics on the connection to BH quantum aspects involving $\hbar$ to
Appendix A.}
\bea
\tau_{\mathrm{relax}} \geq \frac{T_{\mathrm{therm}}}{4\pi} \ .
\eea
In sum, this connection with BH thermodynamics further supports the
interpretation of $T_{\mathrm{therm}}$ as a thermalization
time and the change of perspective
implicit in the passage from the ``density of states'' perspective in (\ref{e:Density}) to
the ``flux'' one in (\ref{e:Flux})~\footnote{From a BH time evolution perspective,
we can rewrite the first law (\ref{e:BH_1st}),
in a dynamical rather than a variational version
\bea
\frac{d\mathrm{Area}^{\mathrm{Hor}}}{dt} = (32\pi M) \frac{dM}{dt} = T_{\mathrm{therm}} \frac{dM}{dt} \ ,
\eea
that relates this purely QNM thermalization
notion, $T_{\mathrm{therm}}$, with the rate of change of the BH physical quantities.}.
In this ``thermal setting'', the QNM Weyl's law for the Schwarzschild BH
adopts the following form 
\bea
\label{e:BH_QNM_Weyl_v3}
N(\omega) &\sim& \frac{1}{c^2}
\left(\frac{1}{6\pi^2}\right) \Big(T_{\mathrm{therm}} \cdot \mathrm{Area}^{\mathrm{LR}}\Big) \omega^3
+  o(\omega^{2}) \ .\nonumber\\
\eea
as $\omega\to \infty$.
Remarkably, inspired by this thermal setting
(but fully agnostic and independent of thermal issues),
a proper intrinsic geometric expression 
can be formulated for Schwarzschild QNM Weyl's law, as we discuss below.

\bigskip

\subsubsection{Schwarzschild QNM Weyl's law: a geometric expression}
\label{s:a_geometric_expression}
The $(3+1)$-Schwarzschild QNM Weyl's law can be written 
\bea
\label{e:BH_QNM_Weyl_v4}
N(\omega) &\sim&
\frac{1}{c^2} \left(\frac{1}{6\pi^2}\right)  \left(\left(\frac{8\pi}{\kappa}\right) \cdot
\mathrm{Area}^{\mathrm{LR}}\right) \omega^3 +  o(\omega^{2}) \ . \nonumber\\
\eea
This is a genuine geometric expression, depending solely
on BH intrinsic quantities. In particular, it is oblivious to any
heuristic thermal reasoning, though the latter has proved a key catalyst.
Expression (\ref{e:BH_QNM_Weyl_v4}) has the virtue of 
highlighting the relevant mechanisms underlying
the characterization of BH resonances,  namely: `light trapping', on the one hand,
and the time-scale decay of QNM (overtones) and its link to the `BH redshift
effect', on the other hand. More explicitly:
\begin{itemize}
\item[i)] {\em Light trapping}: The factor $\mathrm{Area}^{\mathrm{LR}}$ makes explicit
  the role of the light ring, that is, the sphere characterizing light trapping in the spherically symmetric context.

\item[ii)] {\em QNM time-decay scale and (local) redshift effect}: BH QNM frequencies
  present a horizontal ``band structure'' in the $\omega$-complex upper half-plane, starting
  from the real axis, with width  controlled by the surface gravity
  \bea
  \label{e:Deltaomega_kappa}
  \Delta\mathrm{Im}(\omega_n)\sim \kappa \ .
\eea
There are different manners of looking at this band gap scale structuring the QNM frequency complex
plane~\footnote{We note that Eq.~(\ref{e:Deltaomega_kappa}) entails a `horizontal' band structure
  of QNMs in the $\omega$-complex plane. Interestingly, when considering the ultraviolet instability
  of QNMs~\cite{Jaramillo:2020tuu,Jaramillo:2021tmt,alsheikh:tel-04116011,Gasperin:2021kfv,Jaramillo:2022kuv},
  the BH QNM Weyl's law persists. The key point is that, under such `ultraviolet' perturbations of
  wavenumber $k$,  QNMs in the  $\omega$-complex plane still structure themselves in bands, but the latter
  are no longer horizontal  but `circular' around the origin, with shifted bandwidth $\kappa(k)$
  depending on the wavenumber of the  perturbation. Bands are key, but they do need not to be horizontal.}.
A particularly illuminating one
from a geometric analysis perspective
is the characterisation by Warnick (in the Anti-de Sitter setting)
of the regularity needed to define QNMs \cite{Warnick:2013hba}. Indeed, in order
to define QNMs in a $k$-fold band above the real axis of width $k\cdot \kappa$ (as
proper eigenvalues of a particular non-selfadjoint operator), $H^k$-regularity is
required on the eigenfunctions. That is, the more rapidly decaying the QNMs are the more
regular they are required to be. The key BH geometric property underlying this relation
between the QNM frequency gap (\ref{e:Deltaomega_kappa}) and QNM regularity is the BH redshift
effect~\cite{Dafermos:2005eh}, specifically a {\em local}
redshift effect at the horizon with an exponential (retarded) time-decay controlled
by $\kappa$~\cite{Dafermos:2008en}
(for an astrophysical perspective on the relation between $\kappa$ and redshift in BHs,
see \cite{Zimmerman:2016ajr}).

Interestingly, and `independently' of the redshift effect, it is precisely 
the gap $\Delta\mathrm{Im}(\omega_n)\sim \kappa$ in the imaginary part of
BH QNM asymptotics~\cite{Nollert:1993zz} the key element in the justification by Maggiore~\cite{Maggi08}
of Bekenstein area horizon quantization~\cite{Beken74} (see remarks in appendix~\ref{a:appendix_A}).
  
\end{itemize}

\subsubsection{An alternative geometric expression: effective volume $\mathrm{Vol}_3^{\mathrm{eff}}$}
\label{s:Effective_Volume}
The change of perspective  $[\mathrm{Volume}]\sim[\mathrm{Time}]\cdot [\mathrm{Area}]$,
instead of $[\mathrm{Volume}]\sim[\mathrm{Length}]\cdot [\mathrm{Area}]$ has been key
to get to expression (\ref{e:BH_QNM_Weyl_v4}).
However, making use of Smarr formula~\cite{Smarr:1972kt} (see details in appendix~\ref{a:appendix_B})
an alternative and equivalent expression can be written that permits to identify
an effective $3$-dimensional volume in the BH QNM Weyl's law, completely
characterised in terms of spacetime quantities.

Indeed, using Smarr formula for Schwarzschild we write
\bea
\label{e:T_therm_v2}
\left(\frac{8\pi}{\kappa}\right) = 2 \cdot \frac{\mathrm{Area}^{\mathrm{Hor}}}{M} \ ,
\eea
and we can write (\ref{e:BH_QNM_Weyl_v4}) as
\bea
\label{e:BH_QNM_Weyl_Eff_Volume}
N(\omega) \sim \frac{2}{c^3}
\left(\frac{1}{6\pi^2}\right)  \left(\frac{\mathrm{Area}^{\mathrm{Hor}}
  \cdot \mathrm{Area}^{\mathrm{LR}}}{M}\right)
\omega^3 +  o(\omega^{2})\ , \nonumber\\
\eea
We note (see Appendix~\ref{a:appendix_B}) the fine intertwining
between Euler's theorem in Smarr formula and the BH quantities, leading to the
reintroduction of the factor $2$ in BH QNM Weyl's law~(\ref{e:BH_QNM_Weyl_Eff_Volume}).
Comparing with expression~(\ref{e:BH_QNM_Weyl_v2}) for the Weyl's law, this leads
to the identification of the effective volume $\mathrm{Vol}_3^{\mathrm{eff}}$
\bea
\label{e:Vol_eff_Schwarz_v2}
\mathrm{Vol}_3^{\mathrm{eff}} = \left(\frac{\mathrm{Area}^{\mathrm{Hor}}
  \cdot \mathrm{Area}^{\mathrm{LR}}}{M}\right) \ .
\eea
This is indeed a much more transparent expression than (\ref{e:Vol_eff_Schwarz})
and it provides an effective volume for the BH resonant phenomenon characterised
by purely geometric spacetime quantities associated with the BH horizon and the light-ring
(note that this expression is completely time-slice independent). 
However, it fails to make explicit one of the key effects in expression (\ref{e:BH_QNM_Weyl_v4}),
namely the redshift effect. For this reason, BH QNM Weyl's law~(\ref{e:BH_QNM_Weyl_v4})
constitutes our main proposal, whereas the alternative expression~(\ref{e:BH_QNM_Weyl_Eff_Volume}) 
provides an effective volume.

\subsubsection{Caveats with Schwarzschild's BH QNM Weyl's law}
\label{s:caveats}
Expression~(\ref{e:BH_QNM_Weyl_v4}) for the Schwarzschild BH QNM Weyl's law
presents two potentials problems for its extension to more general BHs:
the vanishing of the surface gravity $\kappa$ when approaching extremality, on the
one hand,
and the loss of a two-dimensional light-ring (sphere) when leaving
spherical symmetry, on the other hand. We comment below on this:
\begin{itemize}
\item[i)] {\em BH QNM Weyl's near extremality}: the vanishing of $\kappa$
  when approaching extremality would lead to a divergence of the QNM counting
  function $N(\omega)$ in this limit, if expression~(\ref{e:BH_QNM_Weyl_v4})
  were still valid in this regime. Remarkably, such a divergence
  in the counting of QNMs is actually found in near-extremal Reissner-Nordstr\"om
  BHs, that have a family of  slowly damped QNMs
  with expressions explicitly given by~\cite{Motl:2003cd,Andersson:2003fh,kim2013quasinormal,zimmerman2016damped,richartz2016quasinormal,cardoso2018j,Daghigh:2024wcl}
  \bea
  \omega_{n, \ell} = i \kappa (n + \ell + 1) \ .
  \eea
  Therefore, when fixing a circle of radius $\omega$ in the complex plane,
  the number of such  slowly damped QNMs contained in this circle diverges
  as $\kappa\to 0$. Such a linear dependence on $\kappa$ in the imaginary part
  of  slowly damped QNMs extends beyond spherical symmetry for Kerr-Newman(-de Sitter)
  BHs (cf. e.g.~\cite{hod2008slow,hod2012resonance,yang2013quasinormal,zimmerman2016damped,hod2018quasinormal}
  and references therein;
  see~\cite{joykutty2022existence,Joykutty:2024ctv} for a rigourous mathematical proof in the Reissner-Nordstr\"om de Sitter
  case)
  \bea
  \label{e:ImOmega_gap}
  \mathrm{Im}(\omega_n) = \kappa \left(n + \frac{1}{2}\right) \ , \ \hbox{ as } \kappa\to 0. 
  \eea
  This behavior of the imaginary part of QNM frequencies near the extremal limit,
  under the further hypothesis that $\mathrm{Re}(\omega_n)$ remains bounded as $n$
  grows, is enough to conclude that inside a given circle of large enough radius
  $\omega$, the number of QNMs frequencies diverges as $\kappa\to 0$ (by accumulation
  along the real axis).
  
  This strongly suggests that the surface gravity factor $\kappa$ in the
  BH QNM Weyl's law~(\ref{e:BH_QNM_Weyl_v4}) and the associated divergence
  of $N(\omega)$ in the BH near-extremal limit~\footnote{Note that the QNM spectrum
    in the extremal case differs qualitatively from the near-extremal limit.
    Such limit is singular and we refer here to the limit from sub-extremal side,
  always with  $\kappa>0$.} are robust features of the ultimate  BH QNM Weyl's law.
  
\item[ii)] {\em Light-ring loss}: The light-ring $2$-dimensional structure is unstable under departure
  from spherical symmetry. Indeed, the trapped region ceases to be a sphere in the
  generic case, trapped light rays rather filling a volume. In principle, this fact
  hinders our reasoning for choosing the ``light ring'' prescription
  $\mathrm{Area}^{\mathrm{eff}}$ in Eq.~(\ref{e:area-effr}) and, consequently, the neat separation
  between redshift and trapping effects in BH QNM Weyl's law ~(\ref{e:BH_QNM_Weyl_v4}).
  We address this point in section~\ref{s:Dyatlov-Zworski}, leading to the
  identification of the actual underlying $2$-dimensional structure.

\end{itemize}

\subsection{Relation with Dyatlov \& Zworski BH QNM Weyl's law}
\label{s:Dyatlov-Zworski}
When dwelling in the selfadjoint setting of section~\ref{s:Weyl_law},
specifically when dealing with a quantum mechanical system~\footnote{This
  discussion is of course true for other purely classical selfadjoint problems,
like the wave equations in a cavity subject to conservative boundary conditions,
namely the relevant setting in the black body problem.}
whose dynamics are controlled by a Hamiltonian $\hat{H}$,
the corresponding semiclassical description provides a natural manner
of approaching the asymptotics of the counting function $N(\omega)$.
For concreteness, if considering the (one-particle) dynamics in a bounded domain $D$
of dimension $d$ with Hamiltonian $\hat{H}=-h^2\nabla^2 + V(x)$, the number of
states in the semiclassical limit $h\to 0$ is asymptotically
estimated by the appropriate volume in the phase space of the corresponding classical system.
Specifically, given the (cotangent) phase space $T^*D$, locally
parametrized by $(x,p)$, and the hamiltonian function $H(x,p)= p^2 + V(x)$,
we write
\bea
\label{e:counting_phasespace}
N(\omega) \sim  \frac{1}{(2\pi)^d} \mathrm{Vol}_{T^*D}(H\leq \lambda)
\sim \frac{1}{(2\pi)^d} \int_{H\leq \lambda} \!\!\!d^dx d^dp \ .\nonumber\\
\eea
The separation of the double (multiple) integral in two integrals corresponding
to the the configuration space ($\int d^dx$) and momentum space 
$\int d^dp$ amounts for the two factors $\mathrm{Vol}_d(D)$ and $\mathrm{Vol}_d(B^d_1) \lambda^{d/2}$
in Eq. (\ref{e:counting_function_compact}), respectively.

The same semi-classical strategy can be extended to the resonant
QNM case although, as outlined
in section~\ref{s:QNM_Weyl}, new no-trivial issues emerge.
A pedagogical general overview
is presented in reference~\cite{Zworski99}
(for more details in the semiclassical
approach to QNM Weyl's law, see references therein, in particular~\cite{HelSjo86}).
A concrete issue in stark contrast with the selfadjoint case
concerns the fact that QNM frequencies are now generically
complex and can distribute non-trivially in the complex plane.
In particular, it is not clear that the prescription (\ref{e:counting_function_QNMs}) for the
region in the complex region for counting QNM frequencies, namely a cirle
of radius $\omega$, is the
appropriate one. And, more pragmatically, restrictions on such a circular complex
domain may be needed to obtain sharp results on $N(\omega)$ asymptotics.
This specific point plays a key role of the discussion here below
and is at the core of our final BH QNM Weyl's proposal in section~\ref{s:BH_QNM_proposal}.

\subsubsection{Phase space and the trapped set of null rays}\label{PhaseSpace_TrappedSet}
Such semiclassical approach to QNM (scattering resonances) is adopted
by  Dyatlov \cite{Dyatlov:2013hba} and Dyatlov \& Zworski~\cite{Dyatlov:2013fua}
to study the Weyl's law of Kerr and Kerr-de Sitter BHs, as well as
their stationary regular perturbations.

The key dynamical notion in such analysis is the trapped set of null
geodesics, namely the null geodesics (trajectories) in phase space
that never cross the BH horizon, nor escape to infinity.
This notion extends the light ring (LR) structure that
we have employed in our analysis of section~\ref{s:3-dim_case_analytical},
when discussing the spherically symmetric case. On the one hand, it is key
to emphasize that such trapped set actually lives naturally in the phase space of the spacetime null rays.
Note that in our previous discussion of the light ring (LR), the latter
have been considered as a structure in the spacetime, so momentum variables
did not appear. On the other hand, and as commented in section \ref{s:caveats},
when departing from spherical symmetry, trapped light rays are no
longer confined to two-dimensional spheres (at $r=3M$ in Schwarzschild),
but rather fill a volume in spatial sections of the spacetime.
Remarkably, when looking at the trapped set as an object in the
phase space, the spatial sections are still ``two-dimensional''
in a precise sense.

For the sake of concreteness, we sketch the technical points in  \cite{Dyatlov:2013hba,Dyatlov:2013fua}
of relevance for our discussion. Coordinates are chose so that
spatial $t=\mathrm{const}.$ sections $X$ of the (exterior) of Kerr-de Sitter
BH are parametrised as $X=]r_{\rm h}, r_C[\times \mathbb{S}^2$,
    with $r_{\rm h}$ the coordinate radius of the BH (event) horizon and $r_C$
    the cosmological horizon. The relevant spacetime region is given then
    by $\mathbb{R}\times X$ and the corresponding cotangent bundle $T^*(\mathbb{R}\times X)$
    providing the phase space for the geodesic flow is locally parametrized by
    canonical-pair coordinates
    $(t, r, \theta, \phi, \xi_t, \xi_r, \xi_\theta, \xi_\phi)$.
    The null geodesic flow is given by Hamilton's equations with Hamiltonian
    \bea
    G=g^{ab}\xi_a\xi_b=G_r + G_\theta \ ,
    \eea
    where $G_r$ does not depend on $(\theta, \xi_\theta)$ and $G_\theta$ does not depend on $(r,\xi_r)$
    (cf. \cite{Dyatlov:2013hba,Dyatlov:2013fua}). 
    The trapping of null rays is controlled by the flow of $(r,\xi_r)$ and, specifically, the
    trapped set $K$ is given by null geodesics never reaching 
    $r_{\rm h}$, $r_C$. The set $K\subset  T^*(\mathbb{R}\times X)$ can be characterised as~\cite{Dyatlov:2013hba,Dyatlov:2013fua}
    \bea
    \label{e:trapped_set}
    K = \{ G=0, \xi_r= 0, \partial_rG_r=0, \xi\neq 0\} \ .
    \eea
    The trapped set is a smooth $5$-dimensional submanifold of $T^*(\mathbb{R}\times X)$
    so, in particular, it is not the phase space of any object. However, $t=\mathrm{const}$.
    slices of $K$ provides $4$-dimensional manifolds where the restriction of 
    the canonical symplectic form of $T^*(\mathbb{R}\times X$) is not degenerate.
    In other words, the set
    \bea
    \label{e:K_time_sections}
    K_t =  K \cap \{t=\mathrm{const}\}
    \eea
    is a proper $2$-dimensional symplectic manifold, i.e. a proper $2$-dimensional phase space.
    This is the precise sense in which time sections of the trapped set are ``$2$-dimensional
    objects''.

    \subsubsection{Band structure of (slowest decaying) QNMs}
    A key dynamical element in the discussion of trapping is the local expansion rate of null rays transversal
    to the trapped set, denoted by $\nu$ (namely the Lyapunov exponents of the flow, cf.
    \cite{Dyatlov:2013fua}). Specifically, the minimal and maximal expansion rates
    \bea
    0 < \nu_{\mathrm{min}} \leq \nu_{\mathrm{max}} \ ,
    \eea
    play a very important role in the distribution
    of QNM frequencies in the complex plane. Under appropriate conditions
    (normal hyperbolicity of the trapped set) it holds for all $\epsilon>0$
    \bea
    \mathrm{Im}(\omega_n)> \frac{\nu_{\mathrm{min}} -\epsilon}{2} \ ,
    \eea
    (note the opposite sign convention for $\mathrm{Im}(\omega_n)$ with respect to \cite{Dyatlov:2013fua}).
    In addition,  under the pinching condition   
    \bea
    \label{e:pinching}
    \nu_{\mathrm{max}}< 2 \nu_{\mathrm{min}} \ ,
    \eea
    and a stronger condition (r-normal hyperbolicity) on the trapped set,
    a gap appears between the  band of slowest decaying (fundamental) QNMs and
    faster decaying (overtone) bands.  
    In particular no QNMs appear in the band
    \bea
    \label{e:gap}
     \frac{\nu_{\mathrm{max}} +\epsilon}{2}  < \mathrm{Im}(\omega_n) < \nu_{\mathrm{min}} -\epsilon \ .
     \eea
      Such gap permits to control the fundamental QNMs, the latter being confined
     to the band  $\frac{\nu_{\mathrm{min}} -\epsilon}{2}<  \mathrm{Im}(\omega_n) <
     \frac{\nu_{\mathrm{max}} +\epsilon}{2}$ (cf. e.g. Fig. 6 in~\cite{Dyatlov:2013fua}).
     Remarkably, as we comment below, the counting asymptotics of the fundamental
     QNMs is controlled by the phase space volume of
     trapped set sections $K_t$ in (\ref{e:K_time_sections}), leading
     to Dyatlov and Zworski BH QNM Weyl law. On the contrary, 
    when the pinching condition (\ref{e:pinching}) is not satisfied,
    the  gap (\ref{e:gap}) disappears and the fundamental and overtone
    bands merge, complicating the control of the counting asymptotics.

\subsubsection{Weyl's law for the slowest decaying QNMs}
In the light of the previous discussion, and under the pinching condition
(\ref{e:pinching}) it makes sense to study the Weyl's asymptotics
of fundamental QNMs by defining a counting function that restricts
$N(\omega)$ in (\ref{e:counting_function_QNMs}) to the first band
controlled by the expansion rates. Specifically, Dyatlov and Zworski
define the ``slowest decaying'' QNM counting function $N_{\mathrm{DZ}}(\omega)$
as
\bea
\label{e:counting_function_QNMs_DZ}
N_{\mathrm{DZ}}(\omega) = \# \{&& \omega_n \in \mathbb{C}: \hbox{ such that }
|\omega_n|\leq \omega \nn \\
&& \hbox{ and } \mathrm{Im}(\omega_n) \leq \nu_{\mathrm{min}} - \epsilon \} \ ,
\eea
In this setting, a Weyl's law is rigourously proven for the large-$\omega$ asymptotics of
$N_{\mathrm{DZ}}(\omega)$, namely
\bea
\label{e:asymptotics}
N_{\mathrm{DZ}}(\omega) \sim \frac{1}{(2\pi)^2}\mathrm{Vol}_{K_t}(\xi_t^2\leq 1)\; \omega^2 + o(\omega) \ ,
\eea
that follows the structure of the phase space calculation in expression (\ref{e:counting_phasespace}),
by substituting the phase space $T^*X$ by the symplectic manifold provided by time
sections $K_t$ of the trapped set $K$. Remarkably, the two-dimensional character
fits the exponent $2$ in the asymptotic power-law.
We can conclude that the counting of the ``slowest decaying'' QNMs is controlled
by a $2$-dimensional space structure defined in terms of the trapping 
of null geodesics, namely $K_t$ in (\ref{e:K_time_sections}),
though such $2$-dimensional structure does not correspond in general to a
$2$-surface in spacetime: in spherical symmetry is given by the light ring, but
in the generic case it would correspond to some Lagrangian manifold in $K_t$.

\subsubsection{Reconciling $N(\omega)$ and $N_{\mathrm{DZ}}(\omega)$ asymptotics}
\label{s:summing_on_kappa_bands}
In section \ref{s:a_geometric_expression}, we have shown that the counting
asymptotics of Schwarzschild QNMs follow a cubic power law, $N(\omega)\sim \omega^3$,
whereas Dyatlov and Zworski's Weyl law is quadratic, $N_{\mathrm{DZ}}(\omega) \sim\omega^2$:
how does this reconciliates?

The key lies in the horizontal ``band structure'' of QNMs
in the $\omega$-complex upper half-plane discussed in section
\ref{s:a_geometric_expression}, of width $\Delta\mathrm{Im}(\omega)\sim \kappa$ (cf. (\ref{e:ImOmega_gap})):
whereas $N(\omega)$ counts all QNMs in a circle of radius $\omega$, the function $N_{\mathrm{DZ}}(\omega)$
restricts such counting to only the ``first band'' containing the slowest decaying modes.
The latter indeed follows from the close relation (actually, the proportionality~\cite{yang2013quasinormal})
between Lyapunov's exponents $\nu$ and  surface gravity $\kappa$. In order to get from
the quadratic power $\omega^2$ to the cubic one $\omega^3$ one would need to sum
each  slowest decaying mode over its ``overtone tower''.

This is analogous to the argument leading from the $N_{\ell m}(\omega)\sim \omega$ asymptotics
in section \ref{s:QNM_Weyl_one-dim} resulting from the counting in $n$ of QNM overtones
$\omega_n$ over a fixed $(\ell, m)$ slowest decay QNM,
to the inclusion of all QNMs by summing over angular quantum numbers  $(\ell, m)$ and
leading to  $N(\omega)\sim \omega^3$ in section \ref{s:3-dim_case_analytical}.
In the current setting the order of the counting is reversed, namely one
starts with the angular modes by counting in $N_{\mathrm{DZ}}(\omega)\sim \omega^2$ all the
slowest decaying modes controlled by trapped null rays (note that $K$ in (\ref{e:trapped_set})
is an eminently angular structure) and then  $N(\omega)\sim \omega^3$ would be
recovered by summing over the remaining QNM overtones. Note that it is when counting such `radial' overtone
modes that the surface gravity enters,  $N(\omega)\sim \kappa^{-1}$, as in Eq. (\ref{e:Weyl_Schwarz_one-dim_kappa}).

A main outcome of this discussion is that it suggests an avenue to address the
caveat ii) in section \ref{s:caveats}, concerning the generalisation
of expression (\ref{e:BH_QNM_Weyl_v4}) when loosing the light ring structure.
In short, the sections $K_t$ of the trapped set $K$ would provide the $2$-dimensional
structure generalising the light ring. Indeed, in the non-rotating 
($a=J/M^2=0$) Kerr(-de Sitter) case, the $\partial_r G_r=0$ in (\ref{e:trapped_set})
reduces to $r=3M$, precisely the radius of the light ring (photon sphere).
The expression (\ref{e:BH_QNM_Weyl_v4}) for $N(\omega)$ can then
be formally written as
\bea
\label{e:Nomega_rewriting}
N(\omega)&\sim&
\frac{1}{(2\pi)^3}\left(\left(\frac{8\pi}{\kappa}\right)\cdot \mathrm{Area}^{\mathrm{LR}}\right)\cdot
\left(\frac{4\pi}{3}\omega^3\right)  \\
&\sim& \frac{1}{(2\pi)^3}\left(\left(\frac{8\pi}{\kappa}\right)\cdot\mathrm{Vol}(K_t^{\mathrm{space}, a=0})\right)
\cdot \mathrm{Vol}(B_\omega^3) \nn ,
\eea
where $K_t^{\mathrm{space},a=0}$ denotes the spatial (Lagrangian) sections of $K_t$,
for $a=0$,
and $B_\omega^3$ is the $3$-dimensional Euclidean ball of radius $\omega$. We note, in particular,
the formal structure of $N(\omega)$ as a volume in phase space (cf. in (\ref{e:counting_phasespace})),
factorised in effective ``space cavity'' and a ``momentum sphere'' volumes 
\bea
\label{e:factorisation}
N(\omega)&\sim& \frac{1}{(2\pi)^3}\mathrm{Vol}[\mathrm{Space\;Cavity}]\cdot \mathrm{Vol}[\mathrm{Momentum}(\omega)] \ .\nonumber\\
\eea
In contrast to the light ring, $K_t$ is a robust $2$-dimensional structure
extending beyond spherical symmetry, $a\neq 0$.
Expression (\ref{e:Nomega_rewriting}) for $N(\omega)$ suggests its
extension  to the generic case as~\footnote{Although $K_t$ is a
  robust $2$-dimensional structure well-defined in the generic rotating case,
  it does not need to admit a ``canonical'' factorisation into ``space'' and ``momentum'',
  as in the spherically symmetric case in (\ref{e:Nomega_rewriting}). As a consequence, 
  the geometric numerical coefficients in  (\ref{e:BH_QNM_Weyl_dim-3})
  are not necessarily ``sharp''.}
\bea
\label{e:BH_QNM_Weyl_dim-3}
N(\omega) \sim \frac{1}{(2\pi)^3}
\left(\left(\frac{8\pi}{\kappa}\right) \cdot \mathrm{Vol}_{K_t}(\xi_t^2\leq 1)\right) \omega^3  +  o(\omega^{2}) \ . \nonumber\\
\eea

\subsection{BH QNM Weyl's law: a heuristic proposal}
\label{s:BH_QNM_proposal}
Under the light of the previous discussion, we build on the robustness of the notion of trapped set section $K_t$
to propose an extension of  expressions (\ref{e:BH_QNM_Weyl_v4}) and 
(\ref{e:BH_QNM_Weyl_dim-3})  for the $N(\omega)$ asymptotics of a generic ($d+1$)-dimensional
stationary BH.

\medskip

\noindent{\bf Conjecture (a BH QNM Weyl's law).} {\em The large-$\omega$ asymptotics of the
  QNM counting function $N(\omega)$ for a  generic ($d+1$)-dimensional stationary BH is given by the expression}
\bea
\label{e:BH_QNM_Weyl_dim-d}
N(\omega) \sim \frac{1}{(2\pi)^d}
\left(\left(\frac{8\pi}{\kappa}\right) \cdot \mathrm{Vol}_{K_t}(\xi_t^2\leq 1)\right) \omega^d  +  o(\omega^{d-1}) \ , \nonumber\\
\eea
{\em where $\kappa$ is the BH surface gravity and $K_t$ is the $(d-1)$-dimensional symplectic manifold
  obtained by considering constant time sections of the trapped set $K$ in phase space.}

\medskip

\noindent {\bf Remarks}:

\begin{itemize}
\item[i)] {\em Redshift effect and light-trapping}. The factorisation of the multiplicative
  constant in (\ref{e:BH_QNM_Weyl_dim-d}) encodes, in a purely geometric manner,
  the two main effects controlling the asymptotic counting of QNMs, namely the {\em local}
  redshift effect controlled by $\kappa$ and the trapping of light, with the volume of  time sections
  of the trapped set generalising  the light ring area in expression in (\ref{e:BH_QNM_Weyl_v4}).

\item[ii)] {\em Phase space volume}. A natural question to address in relation to expression
  (\ref{e:BH_QNM_Weyl_dim-d})  would be the identification
of the proper phase space $P\subset T^*(\mathbb{R}\times X)$ and the submanifold $H(\omega)$ whose volume
would account for $N(\omega)$
\bea
\label{e:_phase_space}
N(\omega) \sim \frac{1}{(2\pi)^d} \mathrm{Vol}_{P}(H(\omega)) +  o(\omega^{d-1}) \ ,
\eea
thus generalising the phase space factorisation, in the $3$-dimensional context, in expression (\ref{e:factorisation}).
This question requires a rigorous discussion and is beyond the scope of
the heuristic considerations here presented.

\item[iii)] {\em Effective volume}. An important feature of expression (\ref{e:BH_QNM_Weyl_dim-d}) is that
  its multiplicative factor does not correspond, as in the selfadjoint case (\ref{e:counting_function_compact})
  to the volume of a spatial compact cavity $D$. In the BH QNM Weyl's law this factor is fixed by the
  BH geometry, but it does not correspond to a given spatial section of the spacetime geometry. Actually it
  involves both Lorentzian and symplectic notions. One can always formally introduce
  an effective volume $\mathrm{Vol}_d^{\mathrm{eff}}$, although the latter is not the volume of any ``cavity''.
  Specifically, one can rewrite Eq. (\ref{e:BH_QNM_Weyl_dim-d}) as
  \bea
\label{e:BH_QNM_Weyl_eff_vol}
N(\omega) \sim 2 C_d \mathrm{Vol}_d^{\mathrm{eff}} \omega^d  +  o(\omega^{d-1}) \ ,
\eea
to recover Eq. (\ref{e:BH_QNM_Weyl_v2}), by defining the effective volume
\bea
\label{e:general_eff_vol}
\mathrm{Vol}_d^{\mathrm{eff}} := \frac{1}{2\mathrm{Vol}(B_1^d)}
\left(\left(\frac{8\pi}{\kappa}\right) \cdot \mathrm{Vol}_{K_t}(\xi_t^2\leq 1)\right) \ .
\eea
This generalizes $\mathrm{Vol}_3^{\mathrm{eff}}$ in (\ref{e:Vol_eff_Schwarz}) and (\ref{e:Vol_eff_Schwarz_v2}),
but in the non-spherically symmetric case it is not as transparent as expression (\ref{e:Vol_eff_Schwarz_v2}).
A better understanding of the geometric structure of the phase space in (\ref{e:_phase_space}) would be needed.

\item[iv)] {\em Extremality}. Expression (\ref{e:BH_QNM_Weyl_dim-d}) entails the divergence of the
  BH QNM counting function as $\kappa$ tends to zero. Such a divergence in the limit $\kappa\to 0^+$
  does not imply the divergence of $N(\omega)$ in the exact $\kappa=0$ extremal case.
  As commented in section \ref{s:caveats}, such a divergence is consistent
  with the behaviour  $\displaystyle \mathrm{Im}(\omega_n) \sim \kappa (n+1/2)$, 
  as $\kappa\to 0$, for stationary rotating and charged BHs.
\end{itemize}

\section{Conclusions}
\label{s:conclusions}
We have discussed and extended the Weyl's law for BH QNMs introduced in ref. \cite{Jaramillo:2021tmt},
claiming its consistency with the standard Weyl's law for bound states in selfadjoint problems.
In particular, for $(d+1)$-dimensional BHs we have proposed
a Weyl's law of the standard form $N(\omega)\sim \mathrm{Vol}_d^{\mathrm{eff}}\omega^d$.
On the one hand, it recovers the power-law dependence on the (space) dimension $d$.
On the other hand, regarding the multiplicative factor, we have proposed
that the effective volume  $\mathrm{Vol}_d^{\mathrm{eff}}$ playing the role of the
compact cavity volume in the standard Weyl's law,
is controlled by the  light-trapping properties of the BH.
Specifically,
it factorises as
$\mathrm{Vol}_d^{\mathrm{eff}} \sim \left(8\pi/\kappa\right) \cdot \mathrm{Vol}^{\mathrm{trapped}}_{d-1}$,
the first term being controlled by the (local) redshift effect
and the second  one by the  ($d-1$)-volume of the light trapped set.
We claim this result to hold for generic stationary BHs of all
(asymptotically flat, de Sitter and Anti-de Sitter) asymptotics.

\medskip

In a first stage, following a reasoning combining numerical
and analytical heuristics, we have demonstrated for the $(3+1)$-dimensional
vacuum spherically symmetric (Schwarzschild) BH the relation 
$N(\omega) \sim  (\left(8\pi/\kappa\right) \cdot \mathrm{Area}^{\mathrm{LR}}) \cdot \omega^3$
for the large-$\omega$ asymptotic expression of the QNM counting function $N(\omega)$,
where $\mathrm{Area}^{\mathrm{LR}}$ accounts for the area of the light ring at $r=3M$.
This expression contains already the main structural elements of the generic case.
In particular, the separation of the multiplicative effective
volume $\mathrm{Vol}_3^{\mathrm{eff}}$ into a `radial' part proportional
to the inverse of the surface gravity $\kappa$ and and `angular' part
controlled by area spanned by (angularly) trapped light rays.
Whereas the former (radial) factor accounts for the counting in $n$ of overtones with a
fixed $(\ell, m)$, the latter (angular) factor accounts for the counting when
summing over angular $(\ell, m)$. The sum over each `quantum number'
$(n, \ell, m)$ contributes with a unity to the dimension $d=3$.
In summary, from a methodological perspective,
  this first stage ---focused on the spherically symmetric
  (Schwarzschild) BH--- has acted as a `catalyst' to identify the
essential structures that will be present in the generic BH case.

In this spirit, in a second stage we have extended the  reasoning to 
  generic BHs by abstracting the structural elements found in spherical symmetry.
  Specifically, the emerging picture for  BH QNM distribution,
  eventually leading to conjecture (\ref{e:BH_QNM_Weyl_dim-d}),
  is built  on the following two key and robust remarks:
\begin{itemize}
\item[i)] The $\omega$-complex plane presents a band structure with
bandwidth given by the surface gravity, $\Delta \mathrm{Im}(\omega_n) = \kappa$.
\item[ii)] Dyatlov \& Zworski QNM Weyl's law: the time sections $K_t$ of the trapped set $K$ are $(d-1)$-dimensional
(phase space) structures whose volume counts the
QNMs in a band of width $\kappa$ in the $\omega$-complex plane,
in particular the band of fundamental (slowest decaying) QNMs.
\end{itemize}
In other words, the two basic elements employed in the QNM counting
in the Schwarzschild case are  geometric and indeed robust when passing to generic BHs (with all asymptotics).
Point i) is ultimately justified by the (local) redshift effect discussed
by Dafermos \& Rodnianski \cite{Dafermos:2005eh,Dafermos:2008en} (see also Warnick \cite{Warnick:2013hba}),
whereas point ii) permits to connect with the seminal work by
Dyatlov \& Zworski result~\cite{Dyatlov:2013hba,Dyatlov:2013fua}, claiming its validity  
in the generic BH case. Under the light of the Schwarzschild case the reasoning
for the generic case is now straightforward: when counting QNMs in a circle of radius
$\omega$, point ii) permits to count QNMs in each fixed band of width $\kappa$
(under the pinching condition), providing the power-law $\omega^{d-1}$ and the (angular)
($d-1$)-dimensional (phase-space) volume factor $\mathrm{Vol}_{K_t}(\xi_t^2\leq 1)$, whereas
point i) permits to count the number of bands by dividing by $\kappa$ and
upgrades the power-law $\omega^d$. This results in the final generic
asymptotics $N(\omega)\sim (\left(8\pi/\kappa\right) \cdot \mathrm{Vol}_{K_t}(\xi_t^2\leq 1))\cdot \omega^d$.
Denoting by $\mathrm{Vol}^{\mathrm{trapped}}_{d-1}$ the space part of the phase space volume
$\mathrm{Vol}_{K_t}(\xi_t^2\leq 1)$ we get the proposed form for generic BHs.
In sum, conjecture (\ref{e:BH_QNM_Weyl_dim-d})  can be seen as completing Dyatlov \& Zworski BH QNM Weyl's law
by including overtones.

\subsubsection{Some applications of the BH QNM Weyl's law}
Among the possible implications of the proposed BH QNM Weyl's law,
we comment the following:

\begin{itemize}
\item[i)] {\em Observational access to spacetime dimensionality}. Under the (strong)
  assumption that sufficiently many QNM overtones can be retrieved 
  from gravitational wave signals detected by interferometric gravitational antennae,
  an observational QNM counting function $N^{\mathrm{obs}}(\omega)$ could be
  constructed. Upon verification of the form
  \bea
  \label{e:N_obs}
  N^{\mathrm{obs}}(\omega)\sim \mathrm{Vol}^{\mathrm{eff}}_{\mathrm{obs}} \; \omega^{d_{\mathrm{obs}}} \ ,
  \eea
  it follows, from the BH QNM Weyl's law, the estima-tion $d \approx d_{\mathrm{obs}}$ for the space(time)
  dimensionality. This could provide an observational probe into higher-dimensional
  gravity theories in which gravitational modes propagate into the (higher-dimensional) bulk.

\item[ii)]  {\em Dyatlov \& Zworski spacetime dimensionality $d_{\mathrm{obs}}^{\mathrm{DZ}}$}.
  QNM overtone frequencies
  might be spectrally unstable under high-wavenumber perturbations~\cite{Jaramillo:2020tuu,Jaramillo:2021tmt},
  potentially leading to an essential obstruction to their detection. A more promising
  avenue would then be provided by counting only the slowest decaying QNMs (corresponding to
  the first $\kappa$-width band),
  with much more robust stability properties and perhaps  the only observationally ones accessible.
  In this case, fitting the Dyatlov \& Zworski  BH QNM Weyl's law in Eq.~(\ref{e:asymptotics}) with
  \bea
  N^{\mathrm{obs}}_{\mathrm{slowest-decaying}}(\omega)\sim \omega^{d_{\mathrm{obs}}^{\mathrm{DZ}}} \ , 
  \eea
  the space(time) dimension would be $d\approx  d_{\mathrm{obs}}^{\mathrm{DZ}} +1$.

\item[iii)] {\em Black hole physical parameters}. A key property of Weyl's law is that
  it is oblivious to the details of the resonator, apart from its (effective) volume.
  This property, that is fundamental for its successful application in settings such
  as the black body radiation, strongly limits our capacity to observationally
  probe BH physical parameters by counting QNMs. However, the combination of
  $\mathrm{Vol}^{\mathrm{eff}}_{\mathrm{obs}}$ in (\ref{e:N_obs}) with other observational BH
  probes might provide insight, in particular, into the assessment of extremality, where 
  $\mathrm{Vol}^{\mathrm{eff}}_{\mathrm{obs}}$ should diverge.

\item[iv)] {\em Density of QNM states}. If the BH parameters are a priori known, the
  proposed BH QNM Weyl's law provide a prescription for the spectral density of QNM
  states $\rho(\omega) = dN(\omega)/d\omega$ for large $\omega$'s.
  Such a spectral density could be of use to study thermodynamical aspects of the
  field theory on the boundary in AdS/CFT settings. In another line of reasoning,
  $\rho(\omega)$ can provide the asymptotic ``averaged'' density of QNMs to be employed
  in the construction of appropriate spectral statistics for studying universality
  properties of the BH QNM spectrum (cf.~\cite{Pook-Kolb:2020jlr}
  for a discussion  in this line in the selfadjoint case). Pushing even further the
  speculative tone, $\rho(\omega)$ could play a role in 
  quantum/semiclassical gravity issues potentially involving
  highly damped QNM overtones, such as Hawking radiation or the
  semiclassical description of ``quantum'' BHs (cf. e.g. Appendix~\ref{a:appendix_A}).

\end{itemize}

\subsubsection{Future perspectives}

We can comment the following future research points:
\begin{itemize}

\item[i)] {\em Proof of the conjecture.} In constrast with the Dyatlov \& Zworski BH QNM Weyl's law
  for slowest decaying QNMs,
  that has the status of a theorem (cf. Theorem 3. in~\cite{Dyatlov:2013hba}),
  our proposal is simply a conjecture. A future ultimate goal is  to provide a formal proof of it.

\item[ii)] {\em Numerical assessment in general stationary BHs}.
  Systematic numerical assessment of the proposed BH QNM Weyl's law
  in more general BH scenarios:
  
  \begin{itemize}
  \item[a)] {\em Reissner-Nordstr\"om}. Still in the asymptotically
    flat spherically symmetric case, this provides a non-trivial test
    in particular regarding the dependence of ${\mathrm{Vol}}^{\mathrm{eff}}$
    in $\kappa$ and, very importantly, the divergence of $N(\omega)$ when approaching extremality.

  \item[b)] {\em Kerr-Newman}. The assessment of the conjecture in the non-spherically symmetric case
    is crucial regarding the factorisation of ${\mathrm{Vol}}^{\mathrm{eff}}$
    into the `radial' and `angular' part, when the trapped set does no longer correspond
    to a light-ring sphere.

  \item[c)] {\em General asymptotics: de Sitter and Anti-de Sitter.} The argument leading
    to the BH QNM Weyl's law does not strongly depend on asymptotics at infinity.
    Actually, rather on the contrary,
    lying on de Sitter and Anti-de Sitter asymptotics permits to better connect the argument
    with the existing literature.
    These cases offer therefore a testbed.

  \item[d)] {\em Higher dimensions.} Our reasoning builds on explicit constructions that
    are instanced in the ($3+1$)-dimensional case. However the main conceptual elements,
    namely the $\kappa$-width band structure of the $\omega$-complex plane and the $(d-1)$-dimensional
    structure of the trapped set, are robust in  higher dimensions
    so the extension to the ($d+1$)-dimensional case is natural. 
    This expectation would be clearly backed by the explicit numerical assessement in
    higher-dimensional BHs.

  \item[e)] {\em General perturbations.} The universality of the Weyl's law,
    in particular its characterisation in terms of intrinsic features of the BH geometry, 
    prompts the study of perturbations (scalar, electromagnetic, spinorial, etc.)
    beyond the gravitational case,  aiming at assessing such universality.

  \end{itemize}

\item[iii)]  {\em Ultraviolet QNM overtone instability and BH QNM Weyl's law.} Systematic
  study of the impact  of ultraviolet (high-wavenumber) perturbations on the Weyl's law.
  Such perturbations trigger spectral instabilites in the  
  QNM overtones that migrate to new universal logarithmic (Nollert-Price) QNM branches, while still
  satisfying a Weyl's law~\cite{Jaramillo:2021tmt}.
  We will aim at extending the results in \cite{Jaramillo:2021tmt}, in particular
  assessing the ``phase transition'' in the $\mathrm{Vol}^{\mathrm{eff}}$ factor
  (amounting to a shift in $\kappa$) 
  as the ultraviolet perturbation wavenumber increases.

\end{itemize}

\bigskip

\noindent{\em Acknowledgments.} We would like to thank R\'emi M. Mokdad, Johannes Sj\"ostrand and Claude Warnick.
  This work was supported by the French ``Investissements d'Avenir'' program through
project ISITE-BFC (ANR-15-IDEX-03), the ANR ``Quantum Fields interacting with Geometry'' (QFG) project (ANR-20-CE40-0018-02), 
the EIPHI Graduate School (ANR-17-EURE-0002), 
  the Spanish FIS2017-86497-C2-1 project (with FEDER contribution), STFC grant number ST/V000551/1, the VILLUM Foundation (grant no. VIL37766), the DNRF Chair program (grant no. DNRF162) by the Danish National Research Foundation,  the European Union’s H2020 ERC Advanced Grant “Black holes: gravitational engines of discovery” grant agreement no. Gravitas–101052587
   and the European Commission
  Marie Sklodowska-Curie grant No 843152 (Horizon 2020 programme). The authors acknowledge the use of the IRIDIS High Performance Computing Facility, and associated support services at the University of Southampton, in the completion of this work, and CCuB computational resources
  (universit\'e de Bourgogne).

\appendix   

\section{QNM and BH quantization: surface gravity and Weyl's law}
\label{a:appendix_A}
This appendix aims at providing further support to
the structural importance of the ``band structure'' of QNM overtones
in the upper complex plane,
with gap given by $\Delta\mathrm{Im}(\omega_n)\sim \kappa$
(cf. Eq.~(\ref{e:Deltaomega_kappa})). In this case the support comes
from a semiclassical discussion of the origin of Bekenstein's
BH area quantization~\cite{Beken74,Bekenstein2020quantum} in ``quantum gravity'',
namely
\bea
\label{e:Bekenstein}
\Delta A = 8\pi\ell_{\mathrm{Planck}}^2 \ ,
\eea
where $\ell_{\mathrm{Planck}}=\hbar$ (in appropriate units) is the Planck length.
Starting from the first law of black hole mechanics
\bea
\label{e:1st_law_BS_sphersym}
\delta M = \frac{\kappa}{8\pi} \delta A \ ,
\eea
one argues a proportionality between the ``quantum'' of area and mass,
respectively denoted by $\Delta A$ and $\Delta M$, given by  
\bea
\Delta A = \frac{8\pi}{\kappa}\Delta M  \ .
\eea
Then, from the special relativistic relation between mass
and energy and from the quantum mechanical relation $\Delta E = \hbar \omega$
for the frequency of the emitted/absorbed  photon between two energy levels,
a heuristic relation is proposed
\bea
\Delta M \sim \Delta E = \hbar \omega_* \ ,
\eea
for some BH frequency $\omega_*$. Then
\bea
\label{e:Delta_A_omega_*}
\Delta A =  \frac{8\pi\hbar}{\kappa}\omega_*  \ .
\eea
Note the presence of the BH surface gravity $\kappa$. Resorting to
a Bohr's correspondence principle, Hod's conjecture~\cite{hod1998bohr}
proposes to link such $\omega_*$ to 
the large $n$ asymptotic behavior of BH QNMs (cf. 
Eq. (\ref{e:asymptotics_n_Schwarz_QNM}) for the Schwarzschild case).
In particular, taking the real part of the asymptotic large $n$
overtone frequencies, Hod proposes $\omega_* = \ln(3)/(8\pi M)$,
leading to $\Delta A = 4 \ln(3) \ell_{\mathrm{Planck}}^2$.
This alternative to Bekenstein proposal~(\ref{e:Bekenstein}) provides
a tantalizing link between BH QNMs and quantum gravity.
However, as discussed in \cite{Maggi08}, such a prescription
for $\omega_*$ in (\ref{e:Delta_A_omega_*}) suffers
from the serious drawback of leading to non-universal expressions for $\Delta A$,
thus blurring its relation to a would-be universal
quantum gravity notion.

The solution~\cite{Maggi08} proposed to cure this lack of universality lies
precisely on the (universal) band structure of QNMs.
Modelling QNMs by damped harmonic oscillators with
natural (real) frequencies $(\omega_0)_n$,  Maggiore~\cite{Maggi08}
proposed to correct Hod's prescription for $\omega_*$ in terms of such frequencies
$(\omega_0)_n$'s that, in the large-$n$ limit where the Bohr's correspondence principle (should)
apply, behave as $(\omega_0)_n\sim \mathrm{Im}(\omega_n)$. Then, in the  large-$n$
transition $\omega_n\to\omega_{n-1}$,~\cite{Maggi08} proposes
\bea
\label{e:omega_*_Delta_Im}
\omega_*  = (\omega_0)_n - (\omega_0)_{n-1} = \Delta\mathrm{Im}(\omega_n)
\eea
Considering together Eqs.~(\ref{e:Delta_A_omega_*}) and~(\ref{e:omega_*_Delta_Im}),
it is precisely the universal band structure of QNMs, with gap width
$\Delta\mathrm{Im}(\omega_n)\sim \kappa$, that guarantees the universality
of $\Delta A$: not only $\Delta A$ does not depend on elements external to the
BH geometry, but even the very details of BH geometry disappears with the
cancellation of the surface gravity $\kappa$, making of $\Delta A$
a robust candidate for generic `quantum gravity' scenarios.
Moreover, the QNM band gap is precisely the one leading
to Bekenstein's BH area quantization~(\ref{e:Bekenstein}). This is remarkable.

\section{Alternatives expressions for the effective volume}
\label{a:appendix_B}
We give some details of the derivation 
of expression (\ref{e:Vol_eff_Schwarz_v2}) for the effective
volume in section~\ref{s:Effective_Volume}.
First we note that the mass $M$ of a Schwarschild BH
is a homogeneous function of degree $\frac{1}{2}$ in the
area $A$
\bea
M = M(A) = \left(\frac{A}{16\pi}\right)^{\frac{1}{2}} \ .
\eea
From Euler's theorem for homogeneous functions, we write
\bea
\label{e:Euler_sphersym}
\frac{1}{2} M = \left(\frac{\partial M}{\partial A}\right) A \ .
\eea
Expressing the first of law of BHs in spherical
symmetry (\ref{e:1st_law_BS_sphersym}), in the form of an exact differential
$\displaystyle dM = \frac{\kappa}{8\pi}dA$, it follows
\bea
\frac{\partial M}{\partial A} =  \frac{\kappa}{8\pi} \ ,
\eea
leading to Smarr's expression~\cite{Smarr:1972kt} for the BH mass
\bea
\frac{1}{2} M = \left(\frac{\kappa}{8\pi}\right) A \ ,
\ \Leftrightarrow M = 2 \left(\frac{\kappa}{8\pi}\right) A \ .
\eea
We can then write (we note $A=\mathrm{Area}^{\mathrm{Hor}}$)
\bea
\label{e:kappa_Area}
\frac{8\pi}{\kappa} = 2\cdot \frac{\mathrm{Area}^{\mathrm{Hor}}}{M} \ ,
\eea
that upon substitution in (\ref{e:BH_QNM_Weyl_Eff_Volume}) and
identification of factors in (\ref{e:BH_QNM_Weyl_v2}),
with special emphasis on the factor $2$ corresponding to the
two propagating modes,
leads to expression (\ref{e:Vol_eff_Schwarz_v2}) for $\mathrm{Vol}_3^{\mathrm{eff}}$
\bea
\label{e:V_e_eff}
\mathrm{Vol}_3^{\mathrm{eff}} = \left(\frac{\mathrm{Area}^{\mathrm{Hor}}
  \cdot \mathrm{Area}^{\mathrm{LR}}}{M}\right) \ .
\eea
We note the fine intertwining
between the factor $\frac{1}{2}$ in Euler's theorem and BH quantities and the factor $2$
in BH QNM Weyl's law needed to properly identify the effective volume as a
$\mathrm{Vol}_3^{\mathrm{eff}}\sim [\mathrm{Length}]^3$
quantity. If this is a coincidence or it rather encodes some structural information is unclear.  
On the other hand, and in spite of its appeal, expression (\ref{e:V_e_eff}) can be misleading,
since it may suggest the wrong behaviour for $N(\omega)$ when leaving the spherical
symmetry, in particular when approaching extremality.
In this sense, we prefer to keep the expression (\ref{e:BH_QNM_Weyl_v4}) in terms of the surface gravity.

For completeness, we give the analogous of (\ref{e:kappa_Area})
for stationary rotating charged  (Kerr-Newman) BHs. The mass
$M$ is again a homogeneous function of degree $\frac{1}{2}$ in the
area $A$, the angular momentum $J$ and the square of the electric charge $Q^2$
\bea
M&=&M(A,J,Q^2) \nn \\
&=& \left(\frac{A}{16\pi} + \frac{4\pi J^2}{A} + \frac{Q^2}{2} + \frac{\pi Q^4}{A} \right)^{\frac{1}{2}} \ .
\eea
Writing now the first law of BHs in differential form
\bea
dM = \frac{\kappa}{8\pi}dA + \Omega dJ + \Phi dQ \ ,
\eea
where $\Omega$ is the horizon angular velocity and $\Phi$ is the electric scalar potential
on the horizon. Then we have
\bea
\frac{\partial M}{\partial A} =  \frac{\kappa}{8\pi} \ , \
\frac{\partial M}{\partial J} = \Omega \ , \
\frac{\partial M}{\partial Q^2} = \frac{1}{2Q}\Phi \ ,
\eea
and from Euler's theorem it follows
\bea
\frac{1}{2}M = \left(\frac{\kappa}{8\pi}\right) A + \Omega J +  \frac{1}{2Q}\Phi Q^2 \ ,
\eea
from which Smarr's formula~\cite{Smarr:1972kt} follows
\bea
M = 2\left(\frac{\kappa}{8\pi}\right) A + 2 \Omega J + \Phi Q \ ,
\eea
and the expression (\ref{e:kappa_Area}) is extended to
\bea
\frac{8\pi}{\kappa} = 2 \cdot \frac{A}{M- 2 \Omega J - \Phi Q } \ .
\eea
To calculate the effective volume $\mathrm{Vol}_3^{\mathrm{eff}}$
one would need to substitute $\mathrm{Area}^{\mathrm{LR}}$ in terms of
$\mathrm{Vol}_{K_t}(\xi_t^2\leq 1)$ in (\ref{e:BH_QNM_Weyl_dim-3}).
Altough this expression now diverges in extremality, as it should,
it is not more transparent than Eq.~(\ref{e:general_eff_vol}) so we prefer
the expression of the effective volume in terms of the surface gravity
$\kappa$.

\bibliographystyle{spmpsci}
\bibliography{Biblio}

\end{document}